# Quieting a noisy antenna reproduces photosynthetic light harvesting spectra

Trevor B. Arp[1,2,†], Jed Kistner-Morris[1,2,†], Vivek Aji[2], Richard Cogdell[3,4*], Rienk van Grondelle[4,5*], Nathaniel M. Gabor[1,2,4*].

[1]Laboratory of Quantum Materials Optoelectronics, University of California, Riverside, CA 92521, USA.

[2]Department of Physics and Astronomy, University of California, Riverside, CA 92521, USA.

[3] Institute of Molecular, Cell, and Systems Biology, College of Medical, Veterinary, and Life Sciences, University of Glasgow, Glasgow G128QQ, United Kingdom.

[4]Canadian Institute for Advanced Research, MaRS Centre West Tower, 661 University Avenue, Toronto, Ontario ON M5G 1M1, Canada.

[5]Department of Physics and Astronomy, Faculty of Sciences, Vrije Universiteit Amsterdam, De Boelelaan 1081, 1081 HV, Amsterdam, Nederland.

*richard.cogdell@glasgow.ac.uk, r.van.grondelle@vu.nl, nathaniel.gabor@ucr.edu

† These authors contributed equally to this work

**Abstract:**

Photosynthesis is remarkable, achieving near unity light harvesting quantum efficiency in spite of dynamic light conditions and noisy physiological environment. Under these adverse conditions, it remains unknown whether there exists a fundamental organizing principle that gives rise to robust photosynthetic light harvesting. Here, we present a noise-canceling network model that relates noisy physiological conditions, power conversion efficiency, and the resulting absorption spectrum of photosynthetic organisms. Taking external light conditions in three distinct niches - full solar exposure, light filtered by oxygenic phototrophs, and under sea water - we derive optimal absorption characteristics for efficient solar power conversion. We show how light harvesting antennae can be finely tuned to maximize power conversion efficiency by minimizing excitation noise, thus providing a unified theoretical basis for the experimentally observed wavelength dependence of light absorption in green plants, purple bacteria, and green sulfur bacteria.

In photosynthesis, light energy harvesting begins with the absorption of sunlight. Photoexcitation energy is rapidly transferred through an antenna network before reaching the reaction center, where charge transfer converts excitation energy into an electrochemical potential gradient across the photosynthetic membrane.[1] Strikingly, even in the presence of dynamic light conditions, rapidly fluctuating molecular structure, and highly intricate energy transfer pathways,[1-5] the light-to-electron conversion process exhibits near unity quantum efficiency. While the delicate interplay of quantum effects with molecular mechanisms of energy management have been explored across highly diverse phototrophs,[6-9] the elementary connection between highly robust light energy harvesting and energetic fluctuations is not established.

Transforming noisy inputs into quiet outputs represents a general design challenge in network architectures including multi-national energy grids,[10-14] auditory and visual neural networks,[15-18] and nanoscale photocells for next generation optoelectronics.[19] While network inputs exhibit statistical fluctuations (e.g., rapid changes of sunlight absorbed by a leaf or solar panel), network outputs may demand a steady rate of information or energy for optimal performance (e.g., constant power from the grid to maintain indoor lighting). Statistical fluctuations - arising from environmental variations and internal processes - fundamentally limit the throughput efficiency of any network. If the flow of energy (power) into a network is significantly larger or smaller than the flow out of the network required to optimally match the output demand, the network must adapt or be structured in such a way as to reduce the sudden over- or under-flow of energy. When the network fails to manage these sudden fluctuations, the results may be remarkable (e.g., photo-oxidative stress in photosynthetic light harvesting or explosive damage to transformers due to fluctuations in the grid).

Figure 1 illustrates our model, which employs generalizations of networks to extract the essential aspects of photosynthetic light harvesting. We begin by constructing a simple network of nodes connected by links, shown schematically in Fig. 1A. The nodes (points at which lines intercept) and links (connecting lines) represent physical objects: excitation energy levels and intermolecular transfer events within the antenna system, respectively. In photosynthesis, light enters the antenna through a large number of pigment molecules, each of which is a member of a small set of distinct molecular species (e.g., chlorophyll a and b). Similarly, we limit our model to consider light entering the network through two classes of absorbing excitation energy levels, depicted in Fig. 1A as nodes $A$ and $B$ with input rates $\mathcal{P}_A$ and $\mathcal{P}_B$. After absorption, excitation energy moves between internal nodes of the antenna network, representing the excitation of intermediate states within the biological antenna complex.[2,8,20] While many pathways through the network may share intermediate links, each unique pathway through the network (colored lines Fig. 1A) is described by effective transfer rates (probabilities) to the singular output. The inclusion of multiple pathways is implemented within these average throughput rates. After passing through the antenna network, energy exits through the output $O$ at a rate $\Omega$.

By analyzing the stochastic flow of excitation energy, we can characterize the antenna network by statistical averages (power throughput) and fluctuations in the rate of energy flow, which we will call noise (see supplement section S1). The power throughput of the antenna system is determined by external light conditions, the absorption characteristics of the absorbing pigment molecules (Fig. 1B), or input nodes, and the molecular dynamics of the network. The antenna inputs are described in the usual way: light absorption by the pigment molecules is characterized by peak widths $w$, separation $\Delta\lambda = |\lambda_B - \lambda_A|$, and the center wavelength (or average distance) between the peaks $\lambda_0$. The solar spectral irradiance (grey line Fig. 1B) - which varies as light propagates through air, the canopy, or sea water - gives the *average* power available within a given range of wavelengths. Choosing the wavelength of an absorption peak simultaneously specifies both the excitation energy and power entering the noisy antenna. While the excitation energy is inversely proportional to wavelength, the absorbed power $\mathcal{P}_A$ or $\mathcal{P}_B$ entering the network is the integrated product of the spectral irradiance and the absorption characteristics of the light harvesting antenna.

Noise in the antenna arises from two main sources: external light conditions and inherent mismatch between inputs and outputs of the network, which may arise due to fast dynamics in the protein structure and corresponding electronic properties. In photosynthesis, an over-powered antenna will produce excess energy that can drive deleterious back-reactions.[21,22] Conversely, a light harvesting network in an under-powered state produces non-optimal output, since the rate of energy transfer out of the network is fixed by electrochemical processes.[23] Over long periods of time, the degree to which the light harvesting network is over- or under-powered is measured by the mean-squared deviation of the total input power (through $\mathcal{P}_A$ and $\mathcal{P}_B$) from the optimal output power at $\Omega$, or more succinctly, the noise (Fig. 1C) (see supplement section S1). Since the absorbed solar power rarely matches exactly the rate of optimal output, the finely tuned network is that which most effectively reduces the antenna noise.

Tuning only the absorption characteristics, our goal is to find a finely tuned network that spends the least amount of time in a state for which the input power is too large or too small compared to the output of the network, thus maximizing the power conversion efficiency (Fig. 1C). Fig. 2, our main result, shows three prototypical photosynthetic antennae - the light harvesting complex (LHC2) of green plants (Fig. 2A), the light harvesting complex (LH2) of purple bacteria (Fig. 2B), and the chlorosome of green sulphur bacteria (Fig. 2C) - and compares their absorption spectrum (Figs. 2D-F) to that predicted by our model (Figs. 2G-I) (see supplement section S2 for full details). To obtain the results of Fig. 2, our model takes as input the local irradiance spectrum, shown as solid grey lines in Figs. 2D-I. Details of internal protein dynamics and the numerous potential electronic pathways through the network are embedded in rates $p_A$ and $p_B$ that couple the inputs of the network $\mathcal{P}_A$ and $\mathcal{P}_B$ to the output $\Omega$: $p_A\mathcal{P}_A + p_B\mathcal{P}_B = \Omega$. Minimizing the variance (noise) of the average distribution $p_A\mathcal{P}_A + p_B\mathcal{P}_B$ then yields the optimal absorption characteristics for noise-cancellation (see supplement sections S1.1 through S1.3 for mathematical details).

The absorption peak positions and spectral separation predicted under light conditions in air, under canopy, or under seawater (colored lines Fig. 2G, H, I, respectively) show striking quantitative agreement with the absorption spectra of these three important phototrophs. Using only the external light spectrum and the linewidth $w$, the predicted peak center position $\lambda_0$ and separation $\Delta\lambda$ reproduce the measured absorption peaks with 98% accuracy on average (Table 1). In the following, we examine the biophysical origins and biological implications of this astonishing correspondence.

To understand our model more deeply, we first identify a striking general feature of photosynthetic organisms: In Fig. 2, the photosynthetic pigments do not absorb at the maximum solar power. Instead, all three phototrophs exhibit pairs of closely spaced peaks in regions where the spectrum shows a steep rate of change with respect to wavelength. Photosynthetic plants look green because their antenna complexes absorb light across the visible spectrum including the blue and red portions yet reflect green wavelengths (Fig. 2D). Purple bacteria are aquatic oxygenic phototrophs.[24] They have adapted to sunlight that is filtered through the canopy of trees and floating oxygenic phototrophs (grey line Fig. 2E, see supplement section S3) and use a light harvesting complex in which bacteriochlorophyll dominates light absorption away from the visible, including green (Fig. 2E). Green sulfur bacteria are a geographically diverse group of bacteria that are adapted to solar light shining through seawater to depths where it is anaerobic.[25]

They do not absorb the peak intensity of this attenuated light spectrum and instead absorb in the region of steepest spectral rate of change.

This remarkable attribute of photosynthetic light harvesting, observed across three prototypical phototrophs, is readily understood within the noisy antenna model. To see this, Fig. 3 shows the behavior of three noise regimes within the antenna network: *over-tuned*, *fine-tuned*, and *poorly tuned*. While the light conditions are identical for all three cases (grey lines Figs. 3A), we can examine how the noise changes with different absorption characteristics (details of this calculation can be found in supplement section S1.4). When the absorbing peaks are spaced too closely (Fig. 3A top), the inherent antenna noise can be strongly reduced, and in the limit that $\mathcal{P}_A = \Omega = \mathcal{P}_B$ there are negligible fluctuations in the rate of energy flow (Fig. 3B top left). This lower bound to the internal noise cannot be reached in natural photosynthetic antennae, where protein dynamics will always drive fluctuations of intermediate excitation energy transfer events. Rather, the over-tuned antenna noise is directly proportional to, and thus dominated by, changes in the varying light spectrum (Fig. 3B top right). As shown in Fig. 3C top, in the presence of random external fluctuations, the distribution of time spent in an over- or under-powered state is flat. In the over-tuned antenna, the average input rarely matches the optimal output.

A poorly tuned antenna (Fig. 3A bottom) is similarly deficient. If the absorbing peaks are well separated, the antenna spends most of the time over- or under- powered. When the power sources $\mathcal{P}_A$ or $\mathcal{P}_B$ are significantly greater or less than the power sink ($\mathcal{P}_A \gg \Omega \gg \mathcal{P}_B$), the noise (as evidenced by a histogram of the excitation energy) in the poorly-tuned antenna becomes broader as the absorbing peaks become more separated (Fig. 3C bottom). When viewed over long times, the poorly tuned antenna spends too little time outputting the optimal power $\Omega$.

The finely tuned antenna absorbs at specific positions on the spectrum that give rise to robust light harvesting even in the presence of both varying light conditions and substantial internal noise. When compared to the over- and under-tuned cases, the finely tuned antenna allows for intermediate internal noise levels (Fig. 3B middle) yet delivers a narrow distribution of power centered at the optimal output $\Omega$ (Fig. 3C middle). Robustness in light harvesting is thus the ability to output - on average - the optimal rate $\Omega$, yet simultaneously allow for internal noise.

To determine the optimal absorption spectrum (Fig. 2G-I) for robust light harvesting, we compute the spectral positions for which the peaks are as close as possible on the light spectrum (favoring reduced internal noise), yet the difference in the absorbed power $\Delta = \mathcal{P}_A - \mathcal{P}_B$ is maximized (supporting robustness against external variations). This condition is equivalent to maximizing the derivative of the light spectrum with respect to wavelength, thus resulting in absorption peaks in regions of steepest slope (see supplementary Section S1.3). The absorption spectra, and thus the excitation transitions, are tuned so that the time averaged sum of input excitation energy is sharply peaked at the output rate (Fig. 3C middle).

Underwater phototrophs provide an excellent natural experiment to test the predictive strength of our model since the solar spectrum is highly variable as a function of depth.[26] Fig. 4A shows the light spectrum at various depths below the seawater surface. The light intensity is attenuated as depth increases, particularly in the red and infrared, due to absorption and scattering in seawater. By comparing the absorption spectra of sub-surface marine phototrophs, such as green sulfur bacteria, to those predicted by quieting a noisy antenna, we can explore whether the natural photosynthetic absorption spectrum matches our model predictions for the relevant phototroph's preferred depth.

From the solar light spectra shown in Fig. 4A we calculate an optimization parameter $\Delta^{op}$ as a function of $\Delta\lambda$ and $\lambda_0$. $\Delta^{op}$ is a function modified from the calculation of $\Delta = \mathcal{P}_A - \mathcal{P}_B$ such that its maxima quiet a noisy antenna (see supplement section S1.3). Fig. 4B shows an example color map of the magnitude of $\Delta^{op}$ at a depth of 1 m and $w = 15$ nm. Two maxima clearly emerge in the color plot near $\lambda_0 = 400$ and 750 nm. These maxima identify the wavelength characteristics of a finely tuned antenna under seawater. By extracting the value of $\Delta\lambda$ and $\lambda_0$ at the maximum in $\Delta$, we obtain the characteristic absorption spectra of the quiet antenna as a function of seawater depth (Figs. 4C–F). Interestingly, we observe that the right-side absorption peaks blueshift as the red part of the spectrum is increasingly attenuated, while the absorption peaks on the left side of the spectral maximum do not change with seawater depth.

Quieting a noisy antenna under 2 m of seawater accurately reproduces the absorption spectrum of green sulfur bacteria. Green sulfur bacteria use bacteriochlorophyll (BChl) as the predominant light absorbing molecules for underwater photosynthesis. We find a close match to the absorption spectra of BChl c and BChl e and the ideal absorption spectra at 2 m below the surface (compared directly in Fig. 2F, I). In particular, the depth dependent long wavelength peaks match with 95% and 98% accuracy (see Table 1). Although highly adaptable, green sulfur bacteria are known to thrive at 1-2 m below the surface.[27] Green sulfur bacteria thrive under precisely the same conditions for which a light harvesting antenna is finely tuned for solar power conversion.

The remarkable degree to which we are able to reproduce photosynthetic absorption spectra is a surprising result, indicating an underlying organizing principle for light energy harvesting systems: Fluctuations fundamentally limit the efficiency of networks and must be avoided. While diverse, phototrophs across many photosynthetic niches may have adapted to build fluctuation-cancelling light harvesting antennae, onto which other active mechanisms for reducing fluctuations can be added. While the connection of our model to natural antenna systems requires detailed quantum models, our framework gives new insight into how extinction coefficients, delocalization lengths, and radiative rates conspire to reduce noise in natural antennae. Moreover, by developing noise-cancelling antennae as a technological foundation, natural and artificial energy harvesting networks - from bacteria thriving near deep sea thermal vents to extended power grids - could be adapted to efficiently convert noisy inputs into robust outputs.

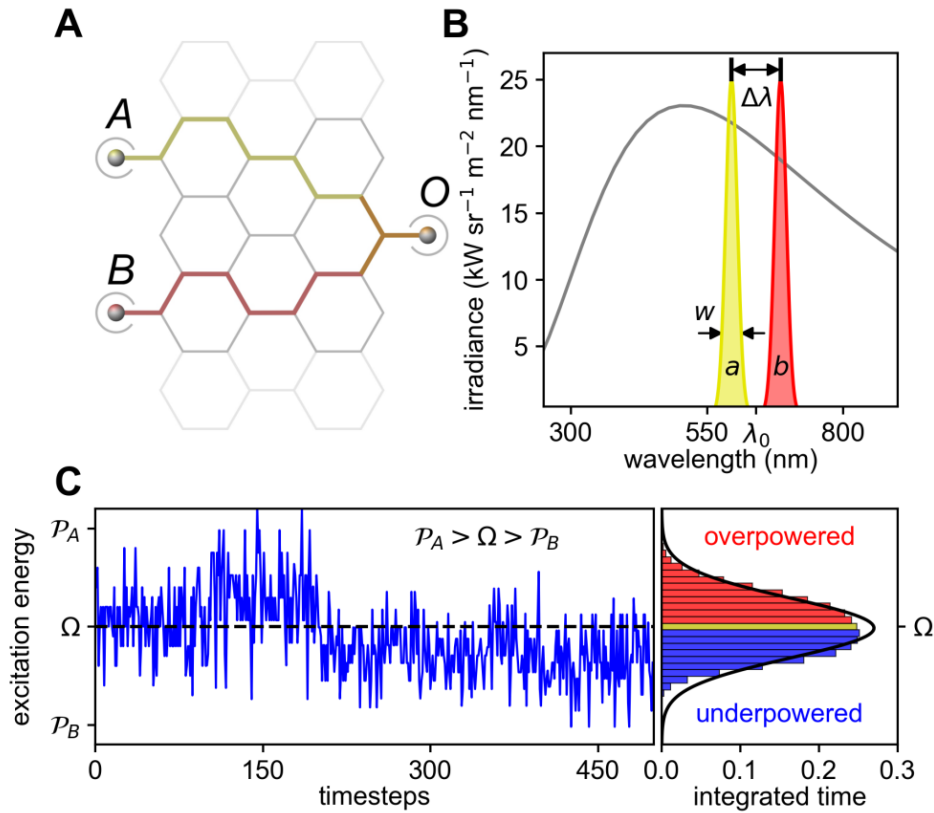

**Fig. 1.** (**A**), Schematic of a photosynthetic antenna reduced into a network with two input nodes $A$ and $B$ with input rates $\mathcal{P}_A$ and $\mathcal{P}_B$, and output $O$ with rate $\Omega$. Energy is absorbed by molecules $a$ and $b$ (at rates $\mathcal{P}_A$ and $\mathcal{P}_B$) and is transferred to the output as usable energy. (**B**), Schematic two-channel antenna absorption spectra (yellow and red) and incident blackbody light source (grey). The quantities $\lambda_0$, $\Delta\lambda$, and $w$ are, respectively, the center wavelength, distance between peaks, and width of the absorption peaks. (**C**), **left,** Simulated average excitation energy as a function of time within a noisy antenna composed of 10 sets of $a$ and $b$ molecules. **right,** Time averaged histogram of the internal energy (detailed in supplement section 1.4). The antenna is subject to internal (fast) and external (slow) fluctuations. Over long timescales the time averaged histogram resembles a normal distribution (black line).

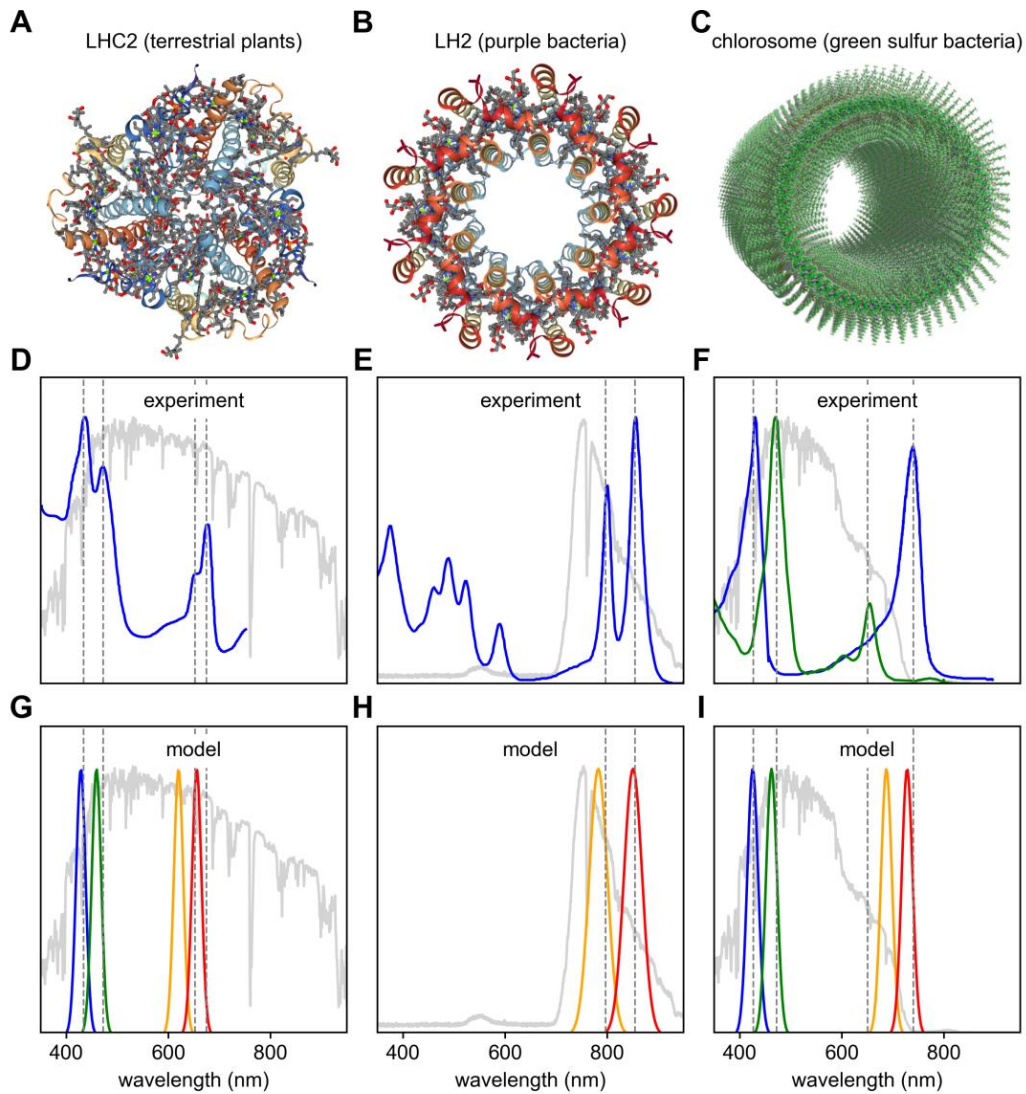

**Fig. 2.** (**A-C**), Molecular structure of the light harvesting antenna LHC2 of green plants, the LH2 of purple bacteria, and the chlorosome of green sulphur bacteria, respectively. (**D**), Absorption spectrum of LHC2 (blue)[31] overlaid on the terrestrial solar spectrum (light grey).[28] (**E**), Absorption spectrum of the LH2 complex overlaid on the solar spectrum below a canopy of leaves (light grey).[32] (**F**), Absorption spectra of bacteriochlorophyll c (blue) and e (green)[33,34] compared to the solar spectrum at 2 m depth of water (light grey).[29,30] (**G-I**), Predicted ideal absorption peaks from optimizing $\Delta = \mathcal{P}_A - \mathcal{P}_B$ for the full solar spectrum, solar spectrum attenuated through canopy, and solar spectrum attenuated through seawater, respectively (see supplement sections S2 and S3 for optimization and spectra details respectively). In (**D-I**), photosynthetic absorption peaks are identified with dashed lines.

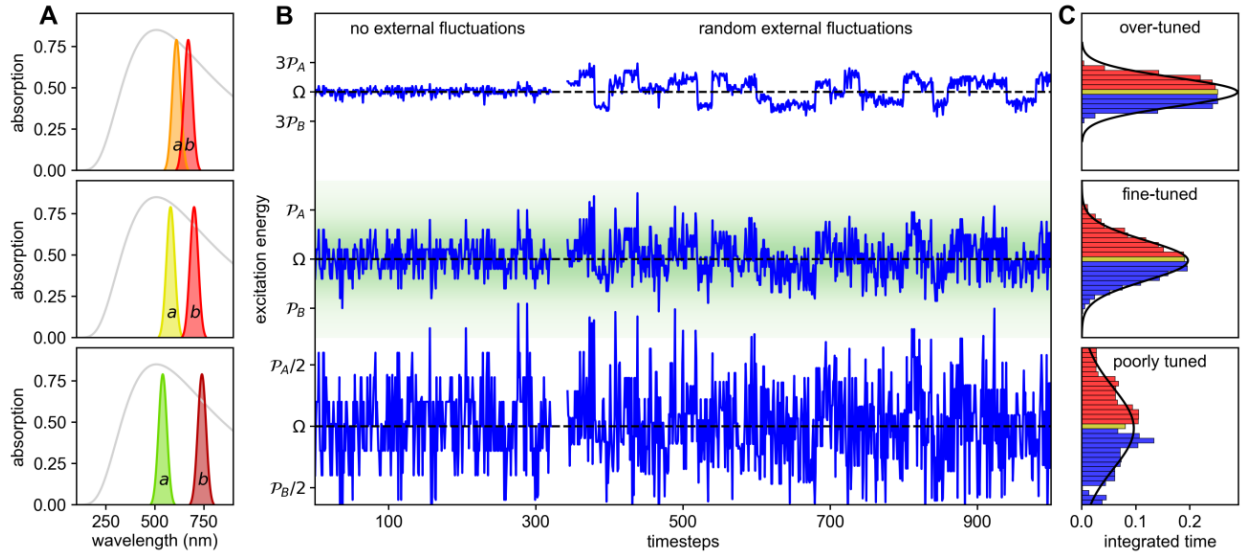

**Fig. 3:** (**A**), Absorption peaks for two absorbers *a* and *b* overlaid on an ideal blackbody solar spectrum ($T = 5500$ K, grey line) for three cases: **top**, two closely spaced absorbers; **middle**, two absorbers separated to optimize the noisy antenna; **bottom**, two widely separated absorbers. (**B**), Simulated excitation energy vs. time for a two-channel antenna with three different values of Δ, comparable to the cases shown in A. Left side shows the excitation energy time traces without external fluctuations. Right side includes random external fluctuations. (**C**), Histograms of time spent in over- (red) and under-powered (blue) states for the three series in B. **top**, the distribution is flat and favors no value. **middle**, the distribution is a sharply peaked normal distribution that favors Ω. **bottom**, the distribution is normal, but wider than in the middle panel.

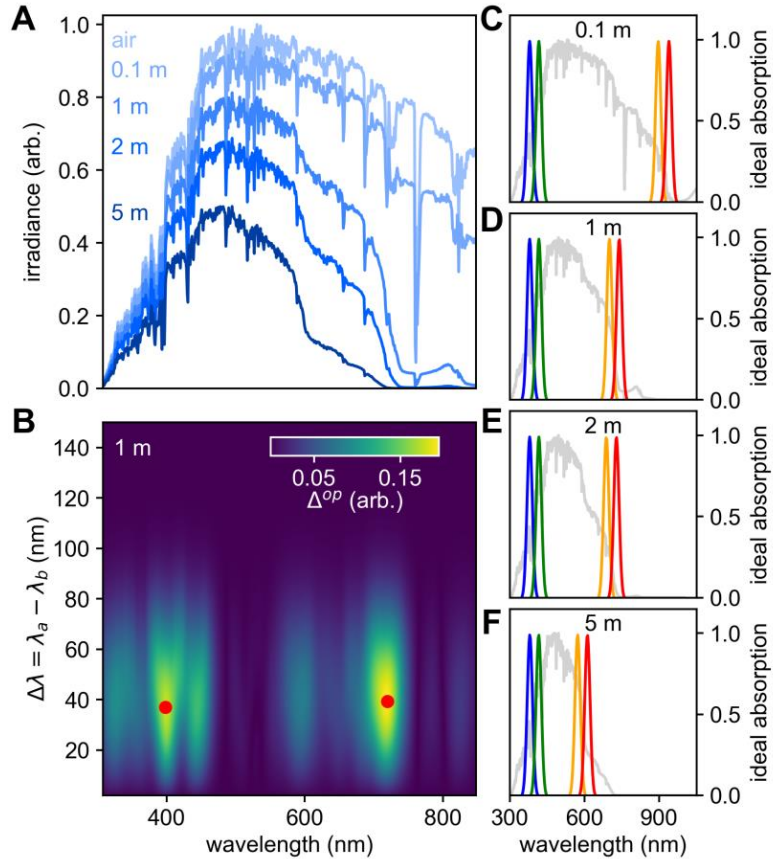

**Fig. 4:** (**A**), Solar spectrum in air and attenuated by various depths of water (labelled).[29,30] (**B**), Optimization landscape calculation of $\Delta^{op}$ versus center wavelength $\lambda_0$ and the peak separation $\Delta\lambda$ for solar spectrum under 1 meter of seawater ($w$ = 15 nm). Red points identify two equally favorable maxima, corresponding to a set of peaks on either side of the spectral maximum. (**C-F**), Ideal absorption peaks predicted from the solar spectrum at each depth. Panel D shows the peaks extracted from the calculation in B, color coded blue, green, orange, red in order to track peak locations with depth.

**Table 1.** Absorption peak data versus model calculation. Comparison between peaks identified from absorption data (Fig. 2 D, E, F) and the absorption peaks of calculated finely tuned light harvesting antennae (Fig. 2G, H, I). The average error is 2.1%.

| Peak Name | Actual Value in nm [eV] | Calculated Value in nm [eV] | Relative % Error | Reference |
|---|---|---|---|---|
| Chlorophyll a 1 | 428 [2.90] | 429 [2.89] | 0.23 [0.34] | *31* |
| Chlorophyll b 1 | 440 [2.82] | 459 [2.70] | 4.32 [4.26] | *31* |
| Chlorophyll b 2 | 652 [1.90] | 620 [2.00] | 4.91 [5.26] | *31* |
| Chlorophyll a 2 | 660 [1.88] | 656 [1.89] | 0.61 [0.53] | *31* |
| LH2 band 1 | 801 [1.55] | 783 [1.58] | 2.25 [1.94] | *32* |
| LH2 band 2 | 857 [1.45] | 851 [1.46] | 0.70 [0.69] | *32* |
| Bacteriochlorophyll c 1 | 431 [2.88] | 426 [2.91] | 1.16 [1.04] | *33* |
| Bacteriochlorophyll e 1 | 461 [2.69] | 462 [2.68] | 0.22 [0.37] | *34* |
| Bacteriochlorophyll e 2 | 655 [1.89] | 688 [1.80] | 5.04 [4.76] | *34* |
| Bacteriochlorophyll c 2 | 740 [1.68] | 728 [1.70] | 1.62 [1.19] | *33* |

**Acknowledgments:** The authors would like to acknowledge valuable discussions with Paul McEuen, Ted Sargent, and Chandra Varma. This work was supported by the Air Force Office of Scientific Research Young Investigator Program (YIP) award no. FA9550-16-1-0216 and through support from the National Science Foundation Division of Materials Research CAREER award no. 1651247. N.M.G. acknowledges support through a Cottrell Scholar Award, and through the Canadian Institute for Advanced Research (CIFAR) Azrieli Global Scholar Award. T.B.A. acknowledges support from the Fellowships and Internships in Extremely Large Data Sets (FIELDS) program, a NASA MUREP Institutional Research Opportunity (MIRO) program, grant number NNX15AP99A. R.J.C. gratefully acknowledges support from the Photosynthetic Antenna Research Center, an Energy Frontier Research Center funded by the U.S. Department of Energy, Office of Science, Office of Basic Energy Sciences under Award Number DE-SC 0001035 and the Biotechnological and Biological Sciences Research Council (BBSRC). R.vG. was supported by the Royal Netherlands Academy of Arts and Sciences and the Canadian Institute for Advanced Research (CIFAR).

**Author contributions:** J. K. M. and T.A. contributed equally to this work, performing detailed analysis and computational modelling. N.M.G. and V.A. conceived the conceptual model, as well as supervised the analytical and computational modelling with additional input from R.J.C. and R.vG. R.J.C., N.M.G. and R.vG. chose which phototrophs should be used as exemplars. All authors contributed to the writing of the manuscript.

**Competing interests:** Authors declare no competing interests.

# Supplementary Materials

**Section S1. The Noisy Antenna Model**

The problem of converting noisy inputs into quiet outputs has relevance in nearly every practical application of network design, ranging from auditory and visual stimulus in neural networks[15-18] to multi-component large scale energy grids.[10-14] Environmental variation (e.g., rapid changes of sunlight in time, varying stimulus impulses in neural networks, etc.) is a considerable obstacle for efficient network operation. If the flow of energy or information *into* a network is significantly larger or smaller than the flow *out of* the network, then the network must continue operation even with the sudden over- or under-flow. When the network fails to manage the over- or under-flow of energy or information, the results can be dramatic (e.g., electrical brownouts are a common example of an intentional or unintentional drop of voltage to some portion of the electrical grid network).

A robust network must produce a stable output even with "noisy" fluctuating inputs. Classical adaptive noise filtering – a technique that utilizes *active* controls and requires an external modulus of control[35-39] – is commonly applied to noisy networks. Such strategies have shown success in various applications from transportation networks[40-45] to electrical networks[46,47], yet these strategies incur an overall cost. In order to maintain active control, additional energy or information must be fed into the network, thus reducing the overall efficiency of energy or information flow.

In this work, we examine the relationship between light harvesting organisms and the light environment, seeking to characterize this relationship using a simplified minimalist model. Such models aim to reduce a complex problem into a form for which calculations become more feasible. A famous example of such a model is that first proposed by Watson and Lovelock[48] to explain global temperature stability in the presence of biofeedback, known as the parable of Daisyworld. Here, we ask whether there exists a simple network topology with *passive* noise reduction, and which requires *no external* assistance to function optimally. Remarkably, we find that such characteristics may emerge in exceedingly simple networks, and we show that carefully routed energy flow within a simple network architecture results in a robust system that inherently quiets internal noise.

The premise of the model we construct is to achieve an optimal tradeoff between minimal noise in energy throughput versus robustness in a noisy environment. To understand why the two are antagonistic consider the case of a single input node A that absorbs at wavelength $\lambda_A$ with power $\mathcal{P}_A$. To minimize throughput noise the absorption rate has to match the output rate $\Omega$, i.e. $\mathcal{P}_A = \Omega$. Such an architecture has no ability to regulate against external fluctuations. Any change in the ambient conditions alters $\mathcal{P}_A$ away from the optimal design. Thus, to gain any ability to adapt the absorber should be at a different power, i.e. $\mathcal{P}_A \neq \Omega$, which in turn introduces noise.

To construct a minimal model, we first consider the case of a single input node. For $\mathcal{P}_A > \Omega$ the absorbing channel switches on and off with probability $p_A$ such that on average input matches output:

$$p_A \mathcal{P}_A = \Omega \ . \tag{1}$$

This randomness gives rise to fluctuations. The input node is *on* some of the time injecting more power than is needed and *off* at other times leaving the network idle. The level of fluctuation is quantified by the standard deviation, which is the square root of the variance. As is conventional, throughout the following, we use the terms variance, fluctuations, and noise interchangeably. The variance, $\sigma^2$ is

$$\sigma^2 = p_A (\mathcal{P}_A - \Omega)^2 + (1 - p_A)\Omega^2 \ . \tag{2}$$

Using Eq. 1 in Eq. 2 simplifies the variance to

$$\frac{\sigma^2}{\Omega^2} = \frac{\mathcal{P}_A}{\Omega} - 1 \ . \tag{3}$$

As anticipated above, any mismatch between input and output power results in noisy throughput. Any external changes that lower $\mathcal{P}_A$ to approach the value of $\Omega$ in turn becomes beneficial ($\sigma^2$ approaches zero), while an increase cannot be regulated against.

Thus, a strategy based on a minimum of two input nodes straddling $\Omega$ is a natural next step. An implicit requirement is that the variations at the two input nodes be correlated. How to satisfy this condition depends sensitively on the nature of the external power spectrum from which the energy is being drawn. We implement the suggested design principles in two steps: 1) we show that two absorbers can reduce noise (Section S1.1, and S1.2) then we calculate the wavelength of the two input nodes that optimizes the tradeoff between noise and robustness for various incident power spectra (Section S1.3).

### S1.1. Two-Channel Noisy Antenna Model

Consider two input nodes at wavelengths $\lambda_A$ and $\lambda_B$ with power $\mathcal{P}_A > \mathcal{P}_B$. We further set $\mathcal{P}_A > \Omega > \mathcal{P}_B$ as suggested by the argument above. At any given time only one of three possibilities occur: 1) power input from node A with probability $p_A$, 2) power input from node B with probability $p_B$, 3) no power absorbed with probability $1 - p_A - p_B$. We explicitly exclude the fourth possibility of power input from both nodes simultaneously as such a process has input power that is much larger than the output, i.e. $\mathcal{P}_A + \mathcal{P}_B \gg \Omega$. This would add large fluctuations without any benefit in robustness due to the fact that the ambient power spectrum has a maximum, thus resulting in an upper bound on large fluctuations above $\Omega$. Given that we are constructing a consistent minimal model that captures the premise outlined above, and the fact that the results are in agreement with observations, provides a validation of this assumption. The matching of input and output power gives the following relationship:

$$p_A \mathcal{P}_A + p_B \mathcal{P}_B = \Omega. \tag{4}$$

The variance is

$$\sigma^2 = p_A (\mathcal{P}_A - \Omega)^2 + p_B (\mathcal{P}_B - \Omega)^2 + (1 - p_A - p_B)\Omega^2. \tag{5}$$

Using Eq. 4 and Eq. 5, the variance simplifies to

$$\frac{\sigma^2}{\Omega^2} = \left(\frac{\mathcal{P}_A}{\Omega} - 1\right) - p_B \frac{\mathcal{P}_B}{\Omega}\left(\frac{\mathcal{P}_A - \mathcal{P}_B}{\Omega}\right). \tag{6}$$

When the second input node is absent, $p_B = 0$, we recover the single node result. Adding a second node reduces noise for a given $\mathcal{P}_A$.

To find an optimal solution, one has to ensure that Eq. 4 is satisfied with the generic constraint that all probabilities must lie between 0 and 1, i.e. $0 \leq p_A \leq 1$, $0 \leq p_B \leq 1$ and $0 \leq p_A + p_B \leq 1$. A consequence of these restrictions is that the optimization process is rather nontrivial, as it is not possible to vary the probabilities and input powers independently. To make further progress we use the inequality $p_A + p_B \leq 1$. Thus

$$p_B \leq 1 - p_A \tag{7}$$

$$\leq 1 - \left(\frac{\Omega - p_B \mathcal{P}_B}{\mathcal{P}_A}\right). \tag{8}$$

Multiplying both sides by $\mathcal{P}_A$ and solving for $p_B$ gives

$$p_B \leq \frac{\mathcal{P}_A - \Omega}{\mathcal{P}_A - \mathcal{P}_B}. \tag{9}$$

Substituting the inequality in Eq. 6, we note that

$$\frac{\sigma^2}{\Omega^2} \geq \left(\frac{\mathcal{P}_A}{\Omega} - 1\right)\left(1 - \frac{\mathcal{P}_B}{\Omega}\right). \tag{10}$$

Since $0 < \mathcal{P}_B < \Omega$, the *least* noise occurs when the input node exactly matches the output and generically goes up as it gets smaller. For a fixed $\mathcal{P}_A$ this is equivalent to the statement that the noise increases as $\Delta = \mathcal{P}_A - \mathcal{P}_B$ increases. Eq. 10 gives a key insight: *An optimized network in an environment with correlated*

*external fluctuations has Δ just large enough to quiet the noisy inputs. Increasing Δ further adds to the internal noise.*

Having established that two input nodes can reduce internal noise compared to one channel and offers robustness against external fluctuations, the next question to address is the correct choice of $\mathcal{P}_A$ and $\mathcal{P}_B$ for a given the ambient spectrum. Before further progress is made, we remark on several important conclusions. From Eq. 10 we can deduce that there is always a residual, which varies continuously with Δ. If the wavelengths $\lambda_A$ and $\lambda_B$ are far apart, the power spectrum in between need not be smooth and the fluctuations at each are in general uncorrelated. On the other hand, if $\lambda_A$ and $\lambda_B$ are close to each other, it is possible to create a robust network that adapts to smooth variation and correlated noise. Thus, one needs to find the region of the power spectrum that provides the largest bandwidth for adaptation (i.e. large Δ) with small $\Delta\lambda = \lambda_A - \lambda_B$. The larger the value of $\Delta = \mathcal{P}_A - \mathcal{P}_B$, the larger the available window for Ω to be within the bounds $\mathcal{P}_A > \Omega > \mathcal{P}_B$ in order to lower noise. It follows therefore that the upper limit of external fluctuations that the network can regulate against is Δ.

### S1.2. Window of Advantage for the Two-Channel Noisy Antenna Model

To understand the design parameters of the model, we must first ask: under what circumstances does the two-channel model give lower noise than the one channel model? At first look, equation 10 seems to imply that the second channel always suppresses the variance when compared to equation 3. But this comparison only holds if you compare a one channel model with input power $\mathcal{P}_A = \mathcal{P}_A^I$ to a two-channel model where the higher input power, $\mathcal{P}_A = \mathcal{P}_A^{II}$, is the same as the one-channel model, i.e. $\mathcal{P}_A^{II} = \mathcal{P}_A^I$. But this is only one possible comparison, how do we meaningfully compare the one-channel and two-channel models? It's always possible to pick a value of $\mathcal{P}_A^I$ that gives a lower variance than any given two-channel model (ignoring whether that value is stable against external fluctuation), so asking if there is a one channel model that is better that a given two-channel model is not useful.

The two-channel model defines an operable range of power $\mathcal{P}_A > \Omega > \mathcal{P}_B$ and attempts to regulate fluctuations within that range. Thus, it is more useful to understand if - for a given one channel model - there is a two-channel operable range that improves the variance. Therefore, to compare the one and two channel models we ask: for a given one channel model defined by $\mathcal{P}_A^I$ and Ω is there a two-channel model with range $\mathcal{P}_B < \mathcal{P}_A^I \leq \mathcal{P}_A^{II}$ that has a lower variance for the same output Ω?

To compare the one and two channel models we examine their respective variances in the parameter space with ordering $\mathcal{P}_B < \Omega < \mathcal{P}_A^I \leq \mathcal{P}_A^{II}$. For a quantitative measure to compare the internal noise of the models we subtract the one channel variance (equation 3) from the two-channel variance (equation 6, re-written in terms of the width of the operable range Δ), giving:

$$\Sigma = \frac{\sigma_{II}^2 - \sigma_I^2}{\Omega^2} = \left(\frac{\mathcal{P}_A^{II}}{\Omega} - 1\right) - p_B \frac{\Delta}{\Omega}\left(\frac{\mathcal{P}_A^{II} - \Delta}{\Omega}\right) - \left(\frac{\mathcal{P}_A^I}{\Omega} - 1\right) \qquad (11)$$

When $\Sigma < 0$ the two-channel model has a lower variance. For a given one channel model, i.e. fixed $\mathcal{P}_A^I$ and $\Omega$, $\Sigma$ varies within the parameter space $(p_B, \mathcal{P}_A^{II}, \Delta)$.

Fig. S1A shows $\Sigma$ for $(\mathcal{P}_A^{II}, \Delta)$ and arbitrarily chosen $p_B = 0.5$, $\mathcal{P}_A^I/\Omega = 1.5$. The yellow region is where $\Sigma \geq 0$ and darker colors indicate $\Sigma < 0$ where the two-channel model is advantageous. However not all of this parameter space is allowed in the two-channel model or satisfies our ordering $\mathcal{P}_B < \Omega < \mathcal{P}_A^I < \mathcal{P}_A^{II}$. Therefore, we introduce constraints to the parameter space of $\Sigma$. Fig. S1B shows the meaningful parameter space, which is bounded by the constraints (shown as colored lines). The first constraints come from ordering, as shown by the horizontal **blue** line, $\mathcal{P}_A^{II}/\Omega > \mathcal{P}_A^I/\Omega$ and as shown by the **orange** line $\mathcal{P}_B < \Omega < \mathcal{P}_A^I$ or equivalently $\Delta/\Omega > \mathcal{P}_A^I/\Omega - 1$. Then we have constraints that come from the two channel definition $\mathcal{P}_A^{II} > \Omega > \mathcal{P}_B$, which requires that $\mathcal{P}_A^{II}/\Omega > \Delta/\Omega$, as shown with the **red** line, and $\Delta/\Omega > \mathcal{P}_A^{II}/\Omega - 1$, as shown with the **green** line. Furthermore, equation 9 limits the probability $p_B$ which translates to the constraint that $\Delta/\Omega \geq (\mathcal{P}_A^{II}/\Omega - 1)/p_B$ as shown by the **magenta** line.

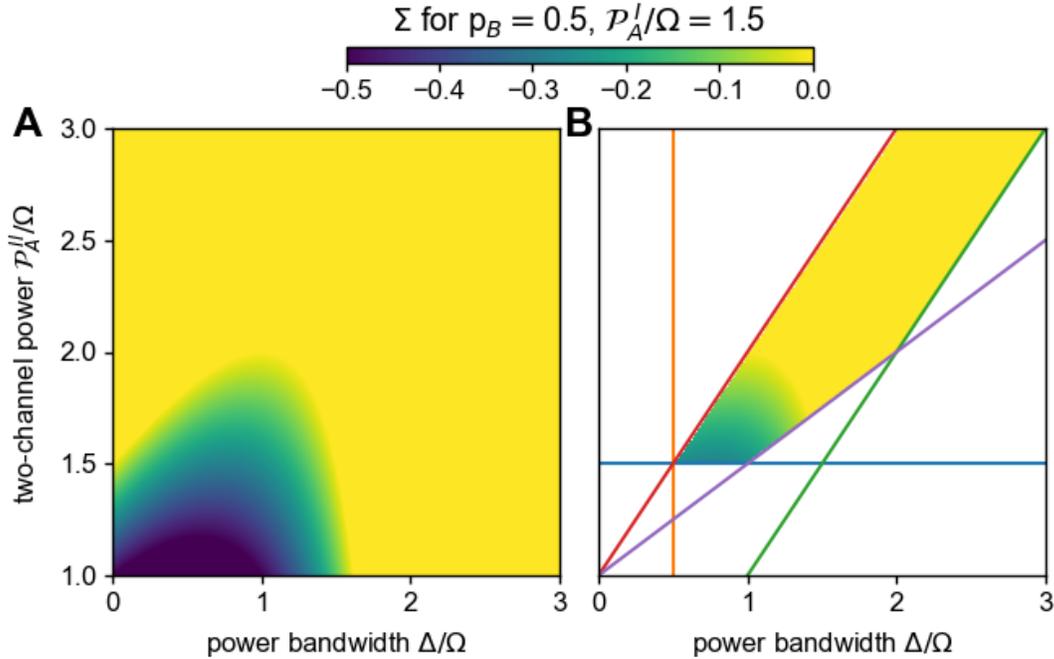

**Fig. S1**: (**A**), Calculation of $\Sigma$ in the unconstrainted parameter space $(\mathcal{P}_A^{II}, \Delta)$ for $p_B = 0.5$ and $\mathcal{P}_A^I/\Omega = 1.5$. Darker colors indicate the two-channel model has a noise advantage over the one channel model. (**B**), The meaningful parameter space of $\Sigma$ that satisfies all the constraints of the model and the ordering, constraints shown as colored lines.

Examining the parameter space of $\Sigma$ reveals that there is a window in which the two-channel model has lower internal noise than the one channel model. Figure S2 explores the full parameter space of $\Sigma$ by showing ($\mathcal{P}_A^{II}$, $\Delta$) for several values of $p_B$ (constant along columns) as a function of various one channel models $\mathcal{P}_A^{I}/\Omega$ (constant along rows). Examining the limits of $p_B$ we observe that for small $p_B \leq 0.1$ there is little advantage, and for $p_B \geq 0.9$ the parameter space that satisfies equation 9 is extremely narrow. This is expected given that either limit the model barely uses one of the channels and is not very different from the one channel model. Looking at moderate range of probability, $0.3 \leq p_B \leq 0.7$ we see that for any given set of parameters there is a window of advantage in which $\Sigma < 0$ and that it generally grows larger as $\mathcal{P}_A^{I}/\Omega$ increases, which is to say, *as the one channel variance increases there is more room to improve.* Therefore, we conclude that if you have a noisy one channel antenna there is a two-channel antenna that will have lower internal noise for the same output. What we do not see in this parameter space is an absolute minimum in which the two-channel model is always best. Rather, for any given one channel model there is a range of possible two channel models which improve upon it.

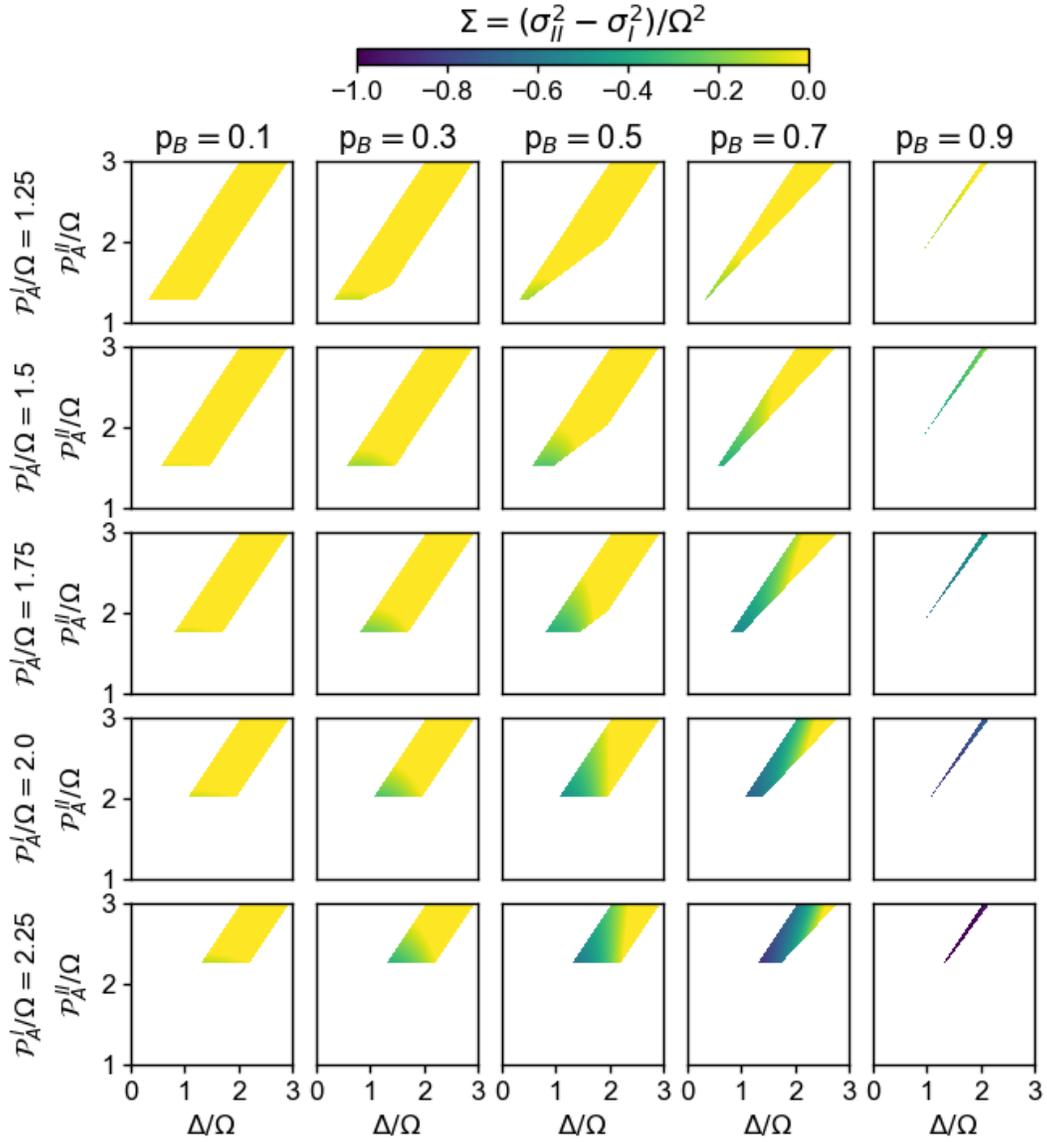

**Fig. S2**: Calculation of $\Sigma$ in the constrained parameter space ($\mathcal{P}_A^{II}$, $\Delta$) for arbitrary values of $p_B$ (constant across columns) and $\mathcal{P}_A^I/\Omega$ (constant across rows).

### S1.3. Calculation of Optimal Absorption Spectrum in the Noisy Antenna

The key conclusion of Section S1.2 is that a two-channel model with a finite window in $\Delta$ is better at suppressing internal noise as compared to a one channel model. On the other hand, to protect against external variability, i.e. the fluctuation in the power spectrum incident on the antenna, one needs $\Delta$ to be as large as possible. Our next step is to determine how these two contradictory properties are best satisfied and develop a quantitative way to find the optimal parameters for a Noisy Antenna in a given light environment. We start with a careful examination of the properties of the power bandwidth $\Delta$.

To determine $\Delta$ and the wavelengths of the input nodes for optimization, a more realistic model for absorption is needed. Unlike the ideal model described above, where all the absorption happens at two fixed wavelengths, absorbers generically operate in a narrow window centered at the $\lambda_A$ and $\lambda_B$ as shown in Fig. 1B. The absorption of the two channels as a function of wavelength, $a(\lambda)$ and $b(\lambda)$, are parameterized by gaussian functions:

$$a(\lambda, \lambda_0, \Delta\lambda, w) = \frac{1}{w\sqrt{2\pi}} \exp\left\{-\frac{[\lambda - (\lambda_0 + \Delta\lambda/2)]^2}{2w^2}\right\} \tag{12}$$

$$b(\lambda, \lambda_0, \Delta\lambda, w) = \frac{1}{w\sqrt{2\pi}} \exp\left\{-\frac{[\lambda - (\lambda_0 - \Delta\lambda/2)]^2}{2w^2}\right\} \tag{13}$$

where $\lambda_0$ is the center wavelength between the two absorbing peaks, $\Delta\lambda$ is the separation between the peaks and $w$ is the width of the peak functions (i.e. the standard deviation). Note that $\lambda_A = \lambda_0 + \Delta\lambda/2$ and $\lambda_B = \lambda_0 - \Delta\lambda/2$. For narrow absorbers, i.e. $w \ll \Delta\lambda$, the precise parametrization - such as Gaussian, Lorentzian, or equivalent - does not change the principal conclusions. The input powers, $\mathcal{P}_A$ and $\mathcal{P}_B$, are obtained in the usual way by integrating the product of the absorption with the irradiance of the solar spectrum, $I(\lambda)$. In other words, $\mathcal{P}_A = \int a(\lambda)I(\lambda)d\lambda$ and $\mathcal{P}_B = \int b(\lambda)I(\lambda)d\lambda$. From these simple definitions, the difference in the absorbed power is given by

$$\Delta(\lambda_0, \Delta\lambda, w) = \int [a(\lambda, \lambda_0, \Delta\lambda, w) - b(\lambda, \lambda_0, \Delta\lambda, w)]I(\lambda)d\lambda. \tag{14}$$

Eq.14 gives a key parameter as $\Delta$ sets the scale of correlated external fluctuations that the noisy antenna is robust against. This suggests a strategy for how to calculate the ideal absorbers for a given solar spectrum.

To insulate from external fluctuations, the first order optimization is a search for the largest value of $\Delta$ in the parameter space $(\lambda_0, \Delta\lambda, w)$. However, this will not quiet a Noisy Antenna as there is a spurious maximum that must be considered, yet discounted. In the case of large $\Delta\lambda$, peak $a$ could sit on the maximum of $I(\lambda)$ and peak $b$ could be on the far edge of $I(\lambda)$. In this scenario $\mathcal{P}_A$ is maximized, and $\mathcal{P}_B$ is effectively zero, thus $\Delta$ is the maximum possible. But this case is clearly outside the window in $\Delta$ where the two-channel model is able to reduce internal noise and in fact is the most extreme version of the *poorly tuned* case discussed in the main text and illustrated in the bottom panels of Fig. 3.

Fortunately, the apparent, but spurious maximum is automatically accounted for when realistic architecture of the network is implemented. The integrated *power* that is absorbed at a given wavelength is transferred to the output node with a finite transition probability, which in turn is proportional to the energy at which the absorption occurs. The energy corresponding to each absorber is $E_{A,B} = hc/\lambda_{A,B}$, where $h$ is Planck's constant and $c$ is the speed of light. If one channel has a significantly larger energy it will be preferred over the other, resulting in the *poorly tuned* case similar to the single input node scenario. Put together our design considerations for the optimization of the Noisy Antenna are as follows:

(1) The two-channel model is advantageous only for a finite range of $\Delta = \mathcal{P}_A^{II} - \mathcal{P}_B^{II}$ within a parameter space defined by $\mathcal{P}_A^{II}$ and $p_B$.

(2) The larger the operable range, $\Delta$, the better the system can protect against external fluctuations.

(3) The typical wavelength of absorption of the two channels should not be too different.

The strategy we adopt to implement these considerations is to determine the operable range for which the absorbers are close in energy ($\Delta\lambda/\lambda_0 \ll 1$) and then determine an optimization function that gives the maximum possible $\Delta$ within this constrained subspace of parameters.

We can estimate the operable range through careful analysis of Eq. 14 with the line shapes specified in Eq. 12 and 13:

$$\Delta(\lambda_0, \Delta\lambda, w) = \int [a(\lambda, \lambda_0, \Delta\lambda, w) - b(\lambda, \lambda_0, \Delta\lambda, w)] I(\lambda) d\lambda \qquad (15)$$

$$= \int \frac{1}{w\sqrt{2\pi}} \left[ \exp\left\{ -\frac{[\lambda - (\lambda_0 + \Delta\lambda/2)]^2}{2w^2} \right\} - \exp\left\{ -\frac{[\lambda - (\lambda_0 - \Delta\lambda/2)]^2}{2w^2} \right\} \right] I(\lambda) d\lambda \qquad (16)$$

$$= \frac{2}{w\sqrt{2\pi}} \exp\left\{ -\frac{\Delta\lambda^2}{8w^2} \right\} \int \exp\left\{ -\frac{(\lambda - \lambda_0)^2}{2w^2} \right\} \sinh\left\{ \frac{\Delta\lambda(\lambda_0 - \lambda)}{2w^2} \right\} I(\lambda) d\lambda. \qquad (17)$$

To make further progress, we invoke empirical, but generic, facts of the spectral irradiance function to evaluate Eq. 17. Specifically, we first recognize that the spectrum $I(\lambda)$ is bounded both in magnitude, with a single maximum, and is limited to a finite window in wavelength. Combining this with the $\exp\{-(\lambda - \lambda_0)^2/2w^2\}$ factor in the integrand of Eq. 17, we conclude that $\Delta(\lambda_0, \Delta\lambda, w)$ is determined by the behavior of the integral in the vicinity of $\lambda_0$. Importantly, expanding Eq. 17 in the vicinity of $\lambda_0$

$$\Delta = \frac{2}{w\sqrt{2\pi}} \exp\left\{-\frac{\Delta\lambda^2}{8w^2}\right\} \int d\lambda \exp\left\{-\frac{(\lambda-\lambda_0)^2}{2w^2}\right\} \sinh\left\{\frac{\Delta\lambda(\lambda_0-\lambda)}{2w^2}\right\}$$
$$\times \left[ I(\lambda_0) + (\lambda-\lambda_0)\frac{dI}{d\lambda}\bigg|_{\lambda_0} + \frac{1}{2}(\lambda-\lambda_0)^2 \frac{d^2I}{d\lambda^2}\bigg|_{\lambda_0} + \cdots \right] \quad (18)$$

we see that all even derivatives in the expansion vanish since sinh is an odd function in $\lambda - \lambda_0$. The leading contribution to the integral comes from the first derivative of $I(\lambda)$. Therefore $\lambda_0$ is in the vicinity of the inflection points of $I(\lambda)$. This yields the intuitive result that maximizing the term puts $\lambda_0$ in the vicinity of the inflection points of $I(\lambda)$.

The natural scale for $\Delta\lambda$ is $2\sqrt{2}w$ appearing in the leading exponential multiplying the integral. While an integration over all $\lambda$ is not very meaningful, given that the expansion is only valid near $\lambda_0$, doing so yields

$$\Delta = \Delta\lambda \sum_{n=0}^{\infty} (2w^2)^n L_n^{\frac{1}{2}}\left(-\frac{\Delta\lambda^2}{8w^2}\right) \frac{d^{2n+1}I}{d\lambda^{2n+1}}\bigg|_{\lambda_0} \quad (19)$$

where $L_n^k(z)$ is the Laguerre Function which are polynomials in $z$ with the highest power going as $n$. For negative z they are also positive definite. Thus, as $n$ increases each successive term adds higher powers of $\frac{\Delta\lambda^2}{8w^2}$ with the sign determined by the *(2n+1)*th derivative evaluated at $\lambda_0$. As anticipated, for $\Delta\lambda < 2\sqrt{2}w$ successive terms become smaller and smaller allowing for a finite value of $\Delta$ consistent with the two channel antenna model. Therefore, we take $\Delta\lambda \leq 2\sqrt{2}w$ as the operable bandwidth that satisfies our design considerations discussed above.

To perform a parameter search for the values of $\lambda_0$ and $\Delta\lambda$ that quiet a Noisy Antenna we want a quantity that satisfies our design considerations, i.e. one maximized in the operable bandwidth while also maximizing power bandwidth within that range. As before we start by considering $\Delta$ as our optimization parameter. Fig. S3A shows two sets of peaks, the blue peaks are when $\Delta\lambda \sim 2\sqrt{2}w$ and the green peaks when $\Delta\lambda$ is the maximum possible i.e. with one peak on the spectral maximum and one on the edge of the spectrum. Evaluating Eq. 14 involves integrating over the peaks, as indicated by the shading, and the result is show in Fig. S3B where $\Delta$ is maximized when $\Delta\lambda$ is the maximum possible. As expected, simply maximizing Eq. 14 will *not* quiet a noisy antenna, but it provides a framework to do so.

To develop a better optimization parameter, we integrate Eq. 14 with modified bounds of integration:

$$\Delta^{\mathrm{op}}(\lambda_0, \Delta\lambda, w) = \int_{\lambda_0-m}^{\lambda_0+m} [a(\lambda, \lambda_0, \Delta\lambda, w) - b(\lambda, \lambda_0, \Delta\lambda, w)] I(\lambda) d\lambda \tag{20}$$

where $m$ is an open parameter that sets a bound to the vicinity of $\lambda_0$. In other words, we are only considering the local contribution to the integral within an interval $2m$ wide, around a point $\lambda_0$. As $\Delta\lambda$ increases, the peaks will fall outside the operable bandwidth and not contribute to the integral. This is shown schematically in Fig. S3C, where only the $2m$ interval, indicated by the shading, is integrated and the wide peaks at $\Delta\lambda^{max}$ are excluded. The result is shown in Fig. S3D where we see that the optimization parameter $\Delta^{op}$ is maximized on the ideal bandwidth, and as $\Delta\lambda \to \Delta\lambda^{max}$ the optimization parameter $\Delta^{op} \to 0$ because green peaks are outside the bounds of integration. The choice of $m$ is somewhat arbitrary, so long as the interval excludes the poorly tuned case and contains the maxima of the peaks when $\Delta\lambda = 2\sqrt{2}w$, then a change in $m$ will not significantly change the location of the maxima. For computational convenience we choose $m = 2w$ without loss of generality, and write

$$\Delta^{\mathrm{op}}(\lambda_0, \Delta\lambda, w) = \int_{\lambda_0-2w}^{\lambda_0+2w} [a(\lambda, \lambda_0, \Delta\lambda, w) - b(\lambda, \lambda_0, \Delta\lambda, w)] I(\lambda) d\lambda \ . \tag{21}$$

Equation 21 is the integral used to calculate all results within this work. Section S2 discusses this optimization for all of the spectra shown in Fig. 2.

From this analysis we can make two basic predictions about the model that can then be verified against real photosynthetic spectra. First, since the leading contribution comes from the first derivative of $I(\lambda)$ we expect $\Delta^{op}$ from Eq. 21 is maximized in the vicinity of the inflection point of the spectrum as a function of $\lambda_0$. We see that all of our optimizations, shown below in section S2, have maxima on the spectral inflection points. In addition, in all prototypical phototrophs shown in main text Fig 2, we find absorption peaks near inflection points in their solar spectra. Second, since we have shown that $\Delta\lambda \sim 2\sqrt{2}w$ is the operable bandwidth, then if the difference between the spectral minimum and maximum is of order $2\sqrt{2}w$ or less there are no optimal peaks because that section of the spectrum is not wide enough. In the case of the purple bacteria, shown in main text Fig. 2E, the left side of the spectrum rises from near zero at 700 nm to the spectral maximum at 750 nm, but this rise takes place over a range less than $2\sqrt{2}w \approx 70$ nm. We subsequently do not observe any peaks in the purple bacteria spectrum on the left side of the spectral maximum, consistent with our model.

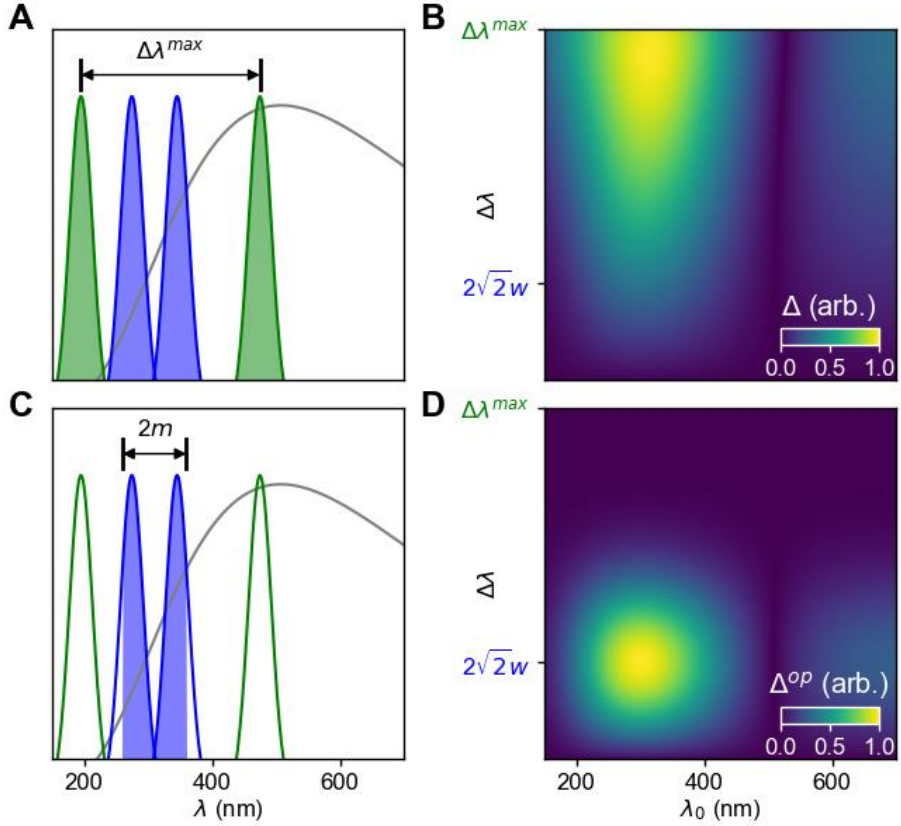

**Fig. S3**: (**A**) Shows two pairs of absorption peaks over an ideal blackbody solar spectrum (grey line). The green peaks have the maximum peak separation, $\Delta\lambda^{max}$ and the blue peaks have the ideal bandwidth $\Delta\lambda = 2\sqrt{2}w$. (**B**) The calculation of Eq. 13 for the ideal blackbody in the parameter space $(\lambda_0, \Delta\lambda)$, as expected $\Delta$ is maximized at $\Delta\lambda^{max}$. (**C**) Shows the absorption peaks again but limits the integrated area to an interval $2m$ wide around $\lambda_0$, indicated by the shading, as in Eq. 18. (**D**) The calculation of Eq. 20 for the ideal blackbody in the parameter space $(\lambda_0, \Delta\lambda)$ with $m = 2w$, which is maximized at $\Delta\lambda = 2\sqrt{2}w$.

### S1.4. Discrete Toy Model

In the main text, we discussed tuning of the noisy antenna in terms of time spent over- and under-powered. To visualize the noise behavior under different choices of input parameters, we calculate the power throughput within a finite system of absorbers as the sum of the absorption events within discrete timesteps. By implementing our model using random trials in the discrete limit, detailed below, we can illustrate how the correct choice of $\Delta$ reduces the sensitivity to external noise while incurring the minimum increase in internal noise. The results of this calculation, shown in Fig. 3B and 3C, provide intuitive visualization for a key statement of our analytical model (made formally in Section S1): An optimized network in an environment with correlated external fluctuations has $\Delta$ just large enough to quiet noisy inputs. Increasing $\Delta$ further adds to the internal noise.

To explore the analytical model of Section S1.1 within a computationally discrete case, we consider a light harvesting network consisting of a small number of absorbing molecules falling into two classes - $a$ and $b$ - that undergo discrete absorption events. This illustrative computation considers a group of 10 absorbing pairs, which was chosen as an order of magnitude estimate within any physically relevant light harvesting network. Our model corresponds to a simplistic network of absorbers with a direct coupling to the output set by a single rate for each of the $a$ or $b$ type absorbers, similar to that shown schematically in Fig. 1. Since the absorbers are simply connected within the network, the number of absorbing pairs sets the noise level.

At each timestep, the molecules will absorb either $\mathcal{P}_A$, with probability $p_a$, absorb $\mathcal{P}_B$ with probability $p_b$, or absorb nothing with probability $1 - p_a - p_b$. The probabilities are set by the equilibrium condition Eq. 4, $p_A \mathcal{P}_A + p_B \mathcal{P}_B = \Omega$ under the symmetric condition $\Omega = (\mathcal{P}_A + \mathcal{P}_B)/2$. There is a range of possible values of $p_a$ and $p_b$ that obey the equilibrium condition. To set the values we define a free parameter $0 \leq \phi \leq 1$ such that $p_B = p_B^{min}\phi + p_B^{max}(1 - \phi)$ where $p_B^{min}$ and $p_B^{max}$ are the minimum and maximum values of $p_B$; from there $p_A$ is set by the equilibrium condition. For the calculations shown in Fig. 3B and 3C, $\phi = 0.05$ was used, but as discussed below, all values of $\phi$ give fundamentally similar results.

The inputs to this calculation are the values of $\mathcal{P}_A$, $\mathcal{P}_B$, and $\Omega$. To simulate external fluctuations in the light environment we add a slowly varying random fluctuation on top of $\mathcal{P}_A$ and $\mathcal{P}_B$, i.e. $\mathcal{P}_A \to \mathcal{P}_A + \mathcal{P}_A \delta P$ and $\mathcal{P}_B \to \mathcal{P}_B + \mathcal{P}_B \delta P$ for some random fluctuation $\delta P$ that changes every 20 timesteps. The output of the calculation is the sum of the energy absorbed from all the absorbing pairs, which gives the excitation energy of the system for that timestep. Due to the equilibrium condition, we expect that the average output should be $10\Omega$, which we see when $\delta P = 0$ (in the figures, $10\Omega$ is re-nomalized to $\Omega$ for simplicity). This calculation simulates excitation energy as a function of time, exhibiting fast stochastic noise and noise due the random external fluctuations. The histogram of this timeseries is what we show in Fig. 3C to illustrate the tuning of the model as a function of $\Delta$.

This highly simplified calculation captures the key behavior of the model regardless of how it is implemented. There are many possible variations. For example, one might consider a system with a different number of absorbing pairs, or with different values of $\phi$. In addition, the model can be made more or less course-grained by considering multiple absorption events occurring per timestep and averaging over them. To increase visual clarity, the calculated data shown in Fig. 3B is averaged over 5 absorption events. The possible variations suggested above do not change the behavior of the model shown in Fig. 3. In all cases when $\Delta$ is small the internal noise of the system is decreased but there is no protection from the external noise, resulting in an energy distribution that mirrors the random external fluctuations. When $\Delta$ is large the external fluctuations away from $\Omega$ are suppressed but the internal noise is large, resulting in a broad distribution of energy. At some intermediate value of Δ external noise without excessive internal noise, resulting in a sharply peaked gaussian energy distribution.

To demonstrate the robustness of the model across many variations, we have developed an interactive visualization of this calculation. This visualization is provided to the reader as a python script called discrete_toy_model.py which is included in our code (see Section S1.5). With it the user may set the parameters (number of absorbers, absorption events per timestep, and $\phi$) of the calculation using command line arguments, then use the slider bar to change the relative value of Δ and see the resulting noise in the timeseries and histogram. Fig. S4 shows screenshots from this tool for three values of Δ, at two different sets of calculation parameters. Though the resulting timeseries are different, the same basic behavior as shown in Fig. 3 emerges. In general, as more absorbers and events are added the less noise is visible in the timeseries and the histogram is clearer.

**S1.5. Noisy Antenna Code**

The code that performs the parameter search of Eq. 21 and the discrete toy model will be released alongside this work when published formally. Code instructions have been removed from this section for the ArXiv version.

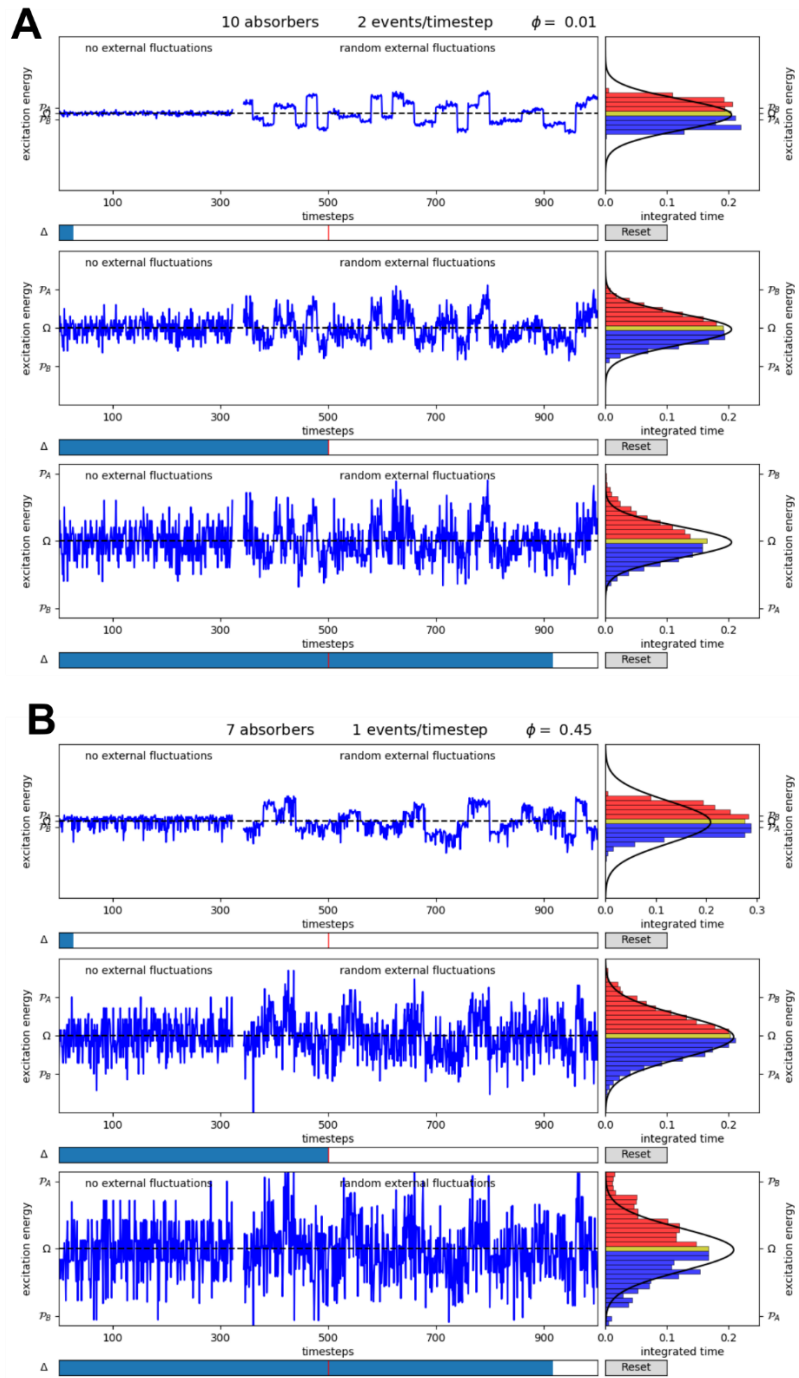

**Fig. S4**. (**A**)**,** Three screenshots of the toy model script, discrete_toy_model.py, for different values of $\Delta$ with 10 absorbers, 2 absorption events per timestep and $\phi = 0.01$. (**B**)**,** The same calculations with 7 absorbers, 1 event per timestep and $\phi = 0.45$.

**Section S2 Ideal Absorption Characteristics from Solar Spectral Data in Distinct Niches**

In order to find the optimum peaks shown in Fig. 2G,H,I we use Eq. 21 to calculate $\Delta^{op}$ for the parameter space defined by $(\lambda_0, \Delta\lambda, w)$ using the solar spectrum in three distinct niches as an input. The solar spectrum is input as $I(\lambda)$, and we find the absorption peaks that correspond to the maxima of $\Delta^{op}$. We first start with the solar spectrum at the surface of the Earth, shown as the grey line in Fig. 2D. Fig. S5A shows the calculation of $\Delta^{op}(\lambda_0, \Delta\lambda, w = 10 \text{ nm})$ for the solar spectrum (low pass filtered to eliminate high frequency spectral noise), and we examine this parameter space to determine the model prediction shown in Fig. 2G.

As discussed in Section S1.3, for a smoothly varying, singly peaked spectrum, like that of a blackbody, there would be two clear maxima at the inflection points on either side of the spectral maximum. However, the complexities of real spectra make this optimization non-trivial. Examining Fig. S5A we see several local maxima on each side of the spectral maximum. This abundance of maxima is due to fine features of the solar spectrum, which absorbers with $w = 13$ nm are too narrow to average over. Fig. S6 performs this calculation for arbitrary $w$. We see that at values of $w > 20$ nm, the calculation is not as sensitive to fine features and shows only two maxima of $\Delta^{op}$, which occur when $\lambda_0$ is near the inflection points on either side of the spectral maximum and the peak separation is $\Delta\lambda \sim 2\sqrt{2}w$.

In photosynthesis, $w$ is fixed by the intrinsic absorption of the absorbing pigment molecule, which is 13 nm for the LHC2. Lacking any way to calculate *a priori* the ideal $w$, we use $w = 13$ nm as an input to our model for chlorophyll and pick out optimal parameters by following the basic model prediction. Therefore, we search for maxima of $\Delta^{op}$ on either side of the spectral maximum, shown in Fig. S5A by the purple and red points. Notably in Fig. S5A we see two bright peaks in $\Delta$ in the 400-450 nm range, these peaks are degenerate for the purpose of our model and we pick the one closer to the spectral maximum, see the discussion in Section S2.2 for details. The purple and red points correspond to two pairs of peaks, which are shown on top of the solar spectrum in Fig. S5B. As expected, the center wavelength of these peaks lies in the neighborhood of the maximum slope of the spectrum, and near the ideal bandwidth $\Delta\lambda \sim 2\sqrt{2}w \sim 37$ nm.

All of the results presented in Fig. 2, and in the following, were calculated by choosing the spectral width of the relevant photosynthetic pigment, and then finding the maximum in $\Delta^{op}(\lambda_0, \Delta\lambda)$. The optimizations for the terrestrial solar spectrum, the solar spectrum under leaves and the solar spectrum under two meters of water are shown in Figs. S5, S7 and S8 respectively. These optimizations generate the main results shown in Fig. 2G,H,I of the main text.

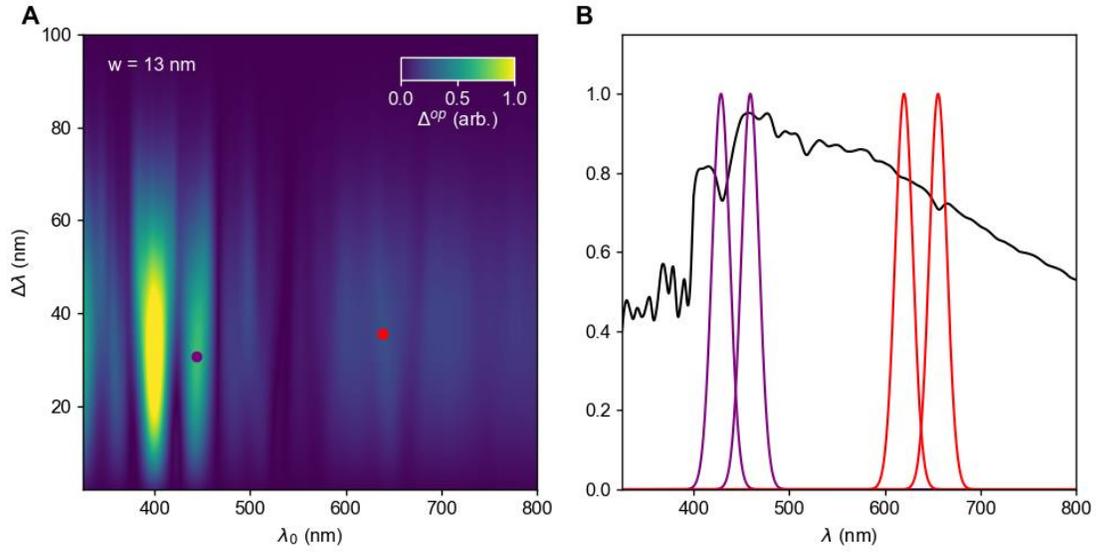

**Fig. S5**: (**A**), Calculation of $\Delta^{op}(\lambda_0, \Delta\lambda, w = 13\text{ nm})$ with the solar spectrum as an input. The red and purple points identify the maxima on either side of the spectra maxima. (**B**), The resulting peak pairs corresponding to these maxima displayed as purple and red lines on top of the filtered solar spectrum (black line).

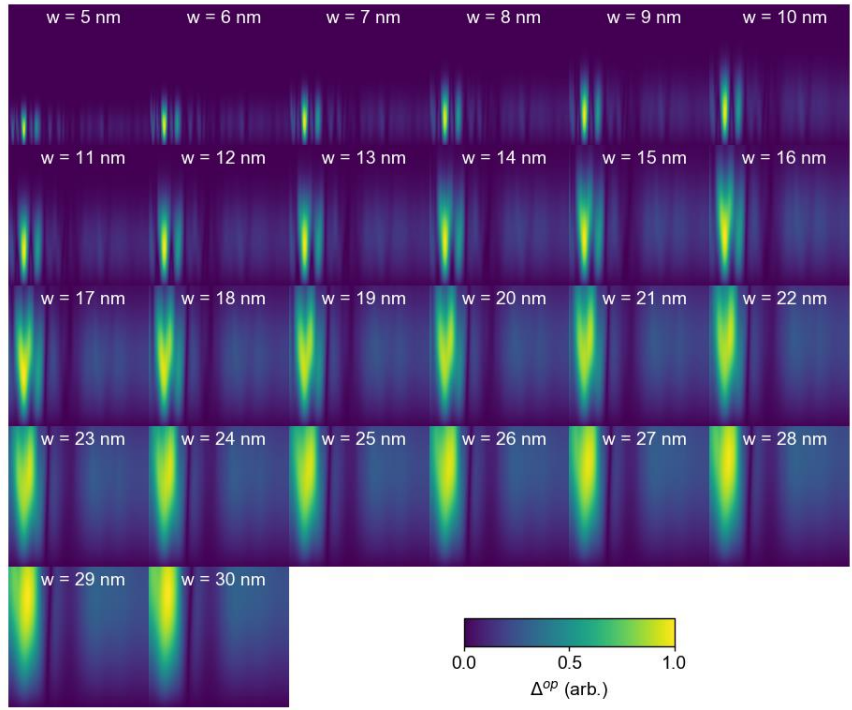

**Fig. S6**. Model optimization of the solar spectrum at the surface of the Earth for variable $w$. The axes of each individual panel are the same as the axes in Fig. S5A.

The characteristic spectrum under canopy (Fig. S7) confirms an interesting feature of the noisy antenna model: for optimal noise cancellation, the peak separation should be $\Delta\lambda \sim 2\sqrt{2}w$. For the optimization shown in Fig. S7 only one pair of peaks is shown, corresponding to the red maximum in Fig. S7A on the right-hand side of the spectral maximum. The maximum on the left-hand side of the spectral maximum is shown as a purple point, however this maximum does not correspond to a *fine-tuned* case, due to the fact that the left-hand side of the spectrum rises too sharply. The left side of the spectrum rises from near zero at 700 nm to the spectral maximum at 750 nm, but this rise takes place over a range less than $2\sqrt{2}w$, the operable bandwidth discussed in section S1.3. Thus, for the optimum peaks on the left side, one peak will be near the maximum, and one near the edge, which is the *poorly tuned* case. Therefore, we do not expect to see a pair of left-side peaks. Indeed, if we look at the absorption spectrum of BChl a (Fig. 2E), we see only a pair of right-side peaks, which correspond well with the peaks found in Fig. S7.

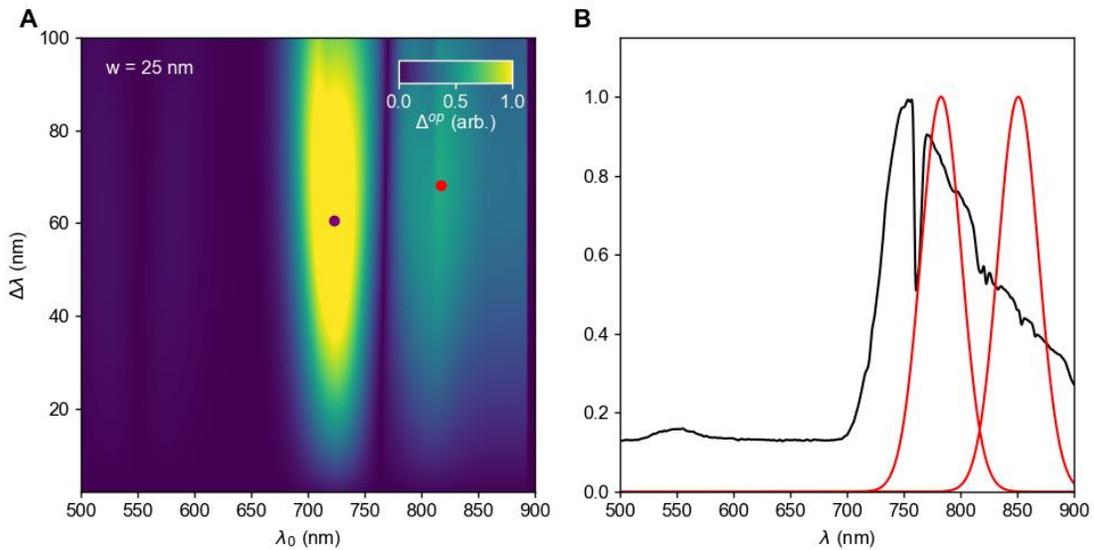

**Fig. S7**. (**A**), Calculation of $\Delta^{op}(\lambda_0, \Delta\lambda, w = 25 \text{ nm})$ with the solar spectrum under cover of leaves (the light environment of Purple Bacteria) as an input. The red and purple points identify the maxima on either side of the spectra maxima. (**B**), The resulting pair of peaks corresponding to the red maxima, on top of the filtered solar spectrum (black line). In this case the purple maximum is disallowed due to the operable bandwidth considerations (see discussion in Section S1.3).

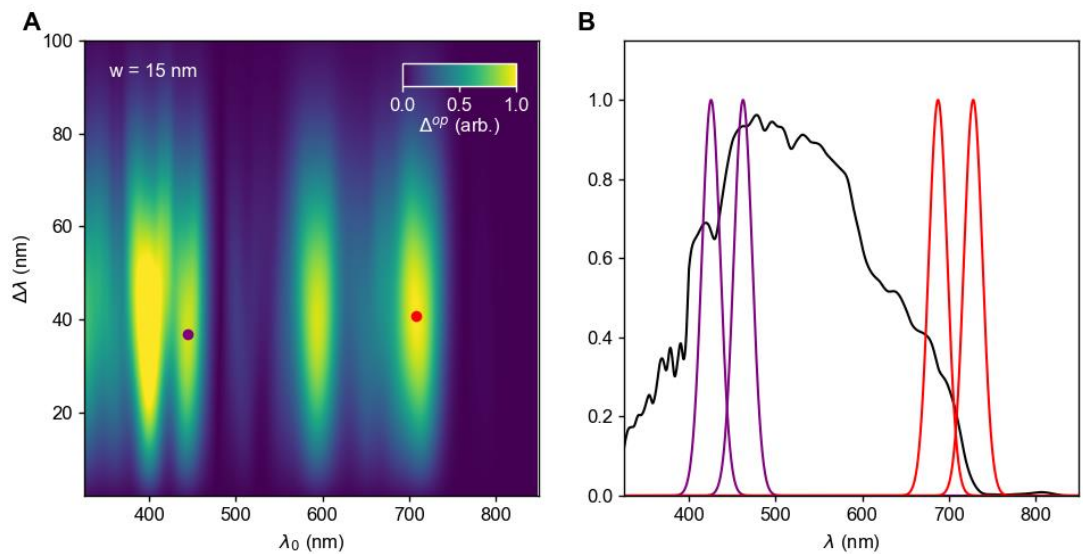

**Fig. S8**. (**A**), Calculation of $\Delta^{op}(\lambda_0, \Delta\lambda, w = 15$ nm) with the solar spectrum two meters underwater (the light environment of Green Sulphur Bacteria) as an input. The red and purple points identify the maxima on either side of the spectra maxima. (**B**), The resulting pairs of peaks corresponding to these maxima displayed as purple and red lines, on top of the filtered solar spectrum (black line).

### S2.1. Comparison of Model with Absorbing Spectra

In main text Fig. 2 we compare the ideal absorption spectra predicted by the model with measurements of the absorption of the light harvesting antenna of actual phototrophic organisms. The absorption measurements shown in Fig. 2D-F are representative of the literature, but here we show that the results exhibit general agreement with multiple absorption spectra reported in the literature. To demonstrate, we will focus on the measurement for terrestrial plants and compare our model prediction to measurements from several sources. Fig. S9A-D show absorption data for the light harvesting complex 2 (LHC2) found in green plants, gathered from various sources.[31,49-51] For all the spectra we identify the four-peak structure with red dots and compare them to the model predictions (grey lines). We also compare the model directly to the absorption spectra of Chlorophyll a and b molecules, from two sources, in Fig. S9E, F.[31,52,53] In all cases we see good agreement with our model's predictions. Fig. S10 shows the predictions (grey dots) versus the actual peaks (red dots) for all measurements, and we see that the error between the predicted and actual peaks is statistical, not systematic, indicating that our model is in general agreement with the literature.

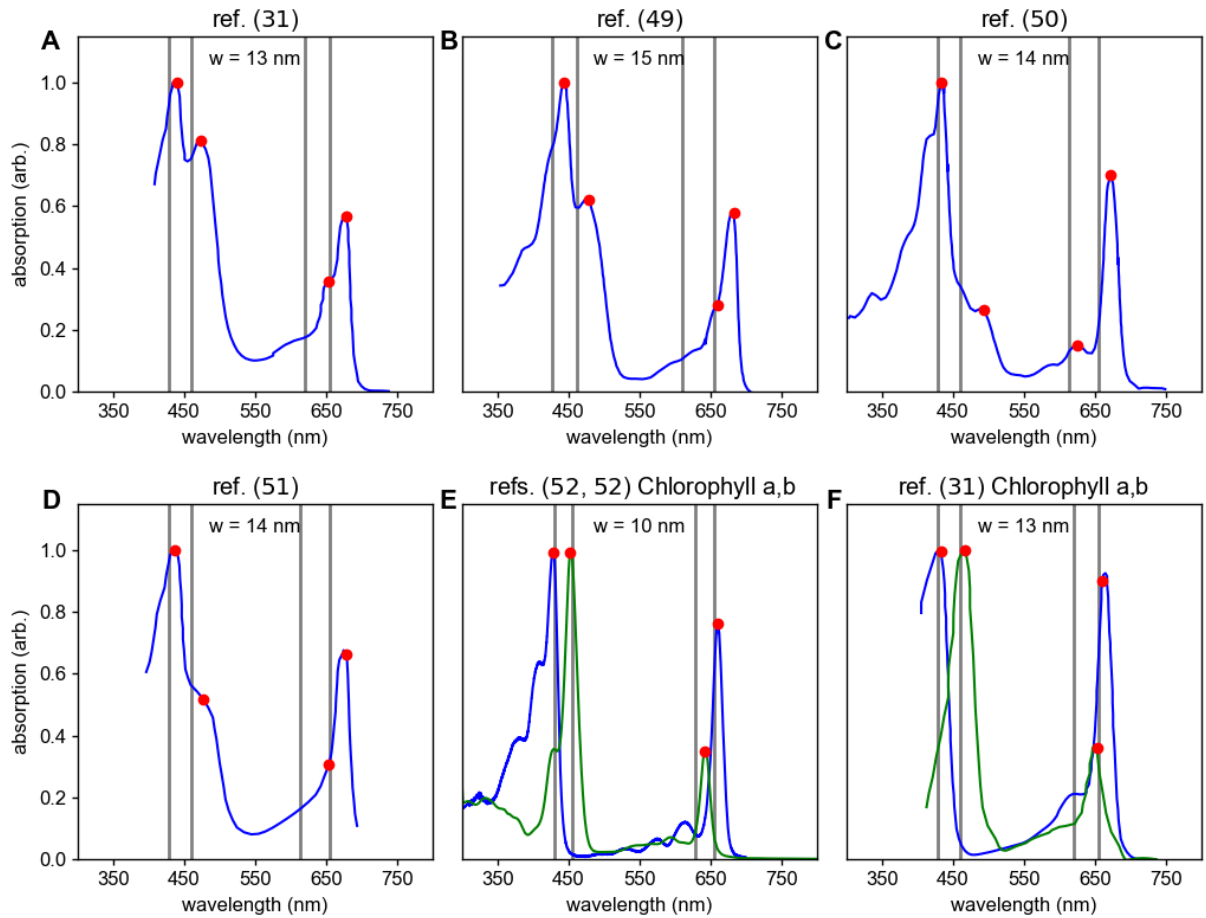

**Fig. S9**. (**A-D**), Data (blue lines) showing the absorbance of Light Harvesting Complex 2 (LHC2) extracted from various sources. Data shown in A is used in main text Fig. 2D. Red dots show maxima in the various absorption spectra and grey lines show the model predictions for a given value of *w*. (**E,F**) Data showing the absorbance of Chlorophyll a (blue line) and b (green line) with identified peaks and model predications.

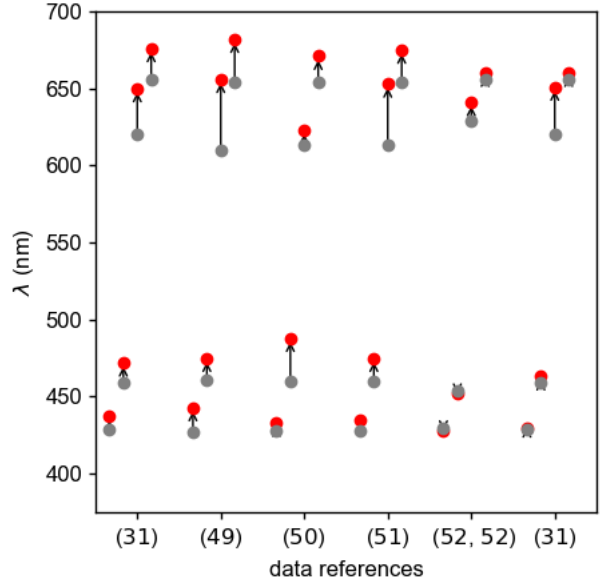

**Fig. S10**. Comparison between the model prediction, grey points, and the absorption peaks identified in Fig. S9, red points. We see good agreement between the model and the data across the literature.

**Section 2.2. Discussion of Degenerate Peaks in the Model Optimization**

In the optimizations for the various solar spectra, shown in Figs. S5, S7 and S8, we see that there are multiple peaks of $\Delta^{op}$, meaning that there are multiple solutions for a given solar spectrum. Furthermore, in the optimization for the solar spectrum (Fig. S5) and underwater (Fig. S8) we do not pick out the largest peaks in $\Delta^{op}$ for the peaks on the left side of the spectral maximum. To explain this, we must examine the peaks on the left side of the solar spectrum in more detail.

Figure S11 shows the model optimization for two slightly different solar spectra. Fig. S11A shows the $\Delta$ calculation for the NREL data of the extraterrestrial solar spectrum.[28] This is the data we use for our terrestrial results to avoid any atmospheric features of the solar spectrum, which are usually variable and would themselves be a source of external fluctuation. Fig. S11B shows the $\Delta^{op}$ calculation for a similar NREL spectrum taken at the surface of the Earth (Direct Circumsolar spectrum form the Air Mass 1.5 measurement). Comparing the spectra (black lines) we see that there are some small differences between the spectra, but they have the same large-scale features. From the colored line traces of $\Delta^{op}$ we see, as in Fig. S6, that at small $w$ there are always two prominent peaks, one near 400 nm and the other near 440 nm, and that as $w$ increases, they merge together into a single peak near 425 nm. In Fig. S11A the 400 nm peak is clearly larger, but in Fig. S11B they are nearly the same for some values of $w$. Thus, there are two clear peaks in this wavelength range, and their relative amplitudes depend on fine features of the solar spectrum that can vary with atmospheric conditions, i.e. the raw amplitudes are not particularly meaningful. In other words, for the purposes of the model these two peaks are degenerate.

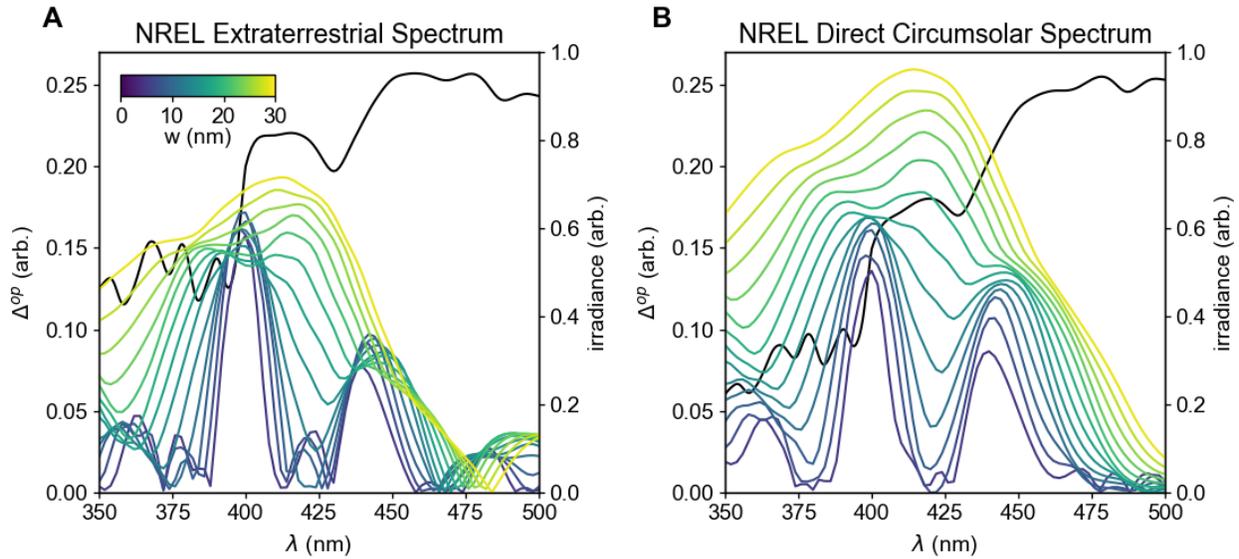

**Fig. S11**. (**A**) The $\Delta^{op}$ optimization for the NREL Extraterrestrial solar spectrum. Colored lines show $\Delta^{op}$ for various values of $w$ at $\Delta\lambda = 2\sqrt{2}w$, compared with the solar spectrum (black line, right axis). (**B**) The $\Delta^{op}$ optimization for NREL Direct Circumsolar solar spectrum, which is attenuated by Earth's atmosphere. Colored lines show $\Delta^{op}$ for various values of $w$ at $\Delta\lambda = 2\sqrt{2}w$, compared with the solar spectrum (black line, right axis).

Given two degenerate peaks in the optimization, how do we choose which solution to include in the model prediction? Given that the two solutions are equal from the perspective of the model, we look at how the peaks line up with the data. Fig. S12 compares the model results for the two degenerate peaks with the LHC2 and Chlorophyll data shown in Fig. S9. Fig. S12A shows the solution corresponding to the $\lambda_0 \sim 400$ nm peak, and we see that it exhibits a large and systematic error when compared with all of the measurements. In contrast, Fig. S12B shows the solution corresponding to the $\lambda_0 \sim 440$ nm peak and see that it lines up well with the measured data, with random error. Therefore, we conclude that nature uses the $\lambda_0 \sim 440$ nm peak. This analysis works equally well when applied to the left side peaks of the underwater spectrum for Green Sulphur Bacteria, which is expected given that the structure of the left side of the solar spectrum is largely unaffected by water (see main text Fig. 4A). There are several potential hypotheses for why the $\lambda_0 \sim 440$ nm peak is selected, the simplest being that if the two peaks are approximately equally advantageous for quieting a noisy antenna nature might select the one with higher power throughput. But testing these hypotheses is beyond the scope of this work. From the perspective of our model there are two degenerate solutions for the left side of the solar spectrum and nature seems to use the one closer to the spectral maximum.

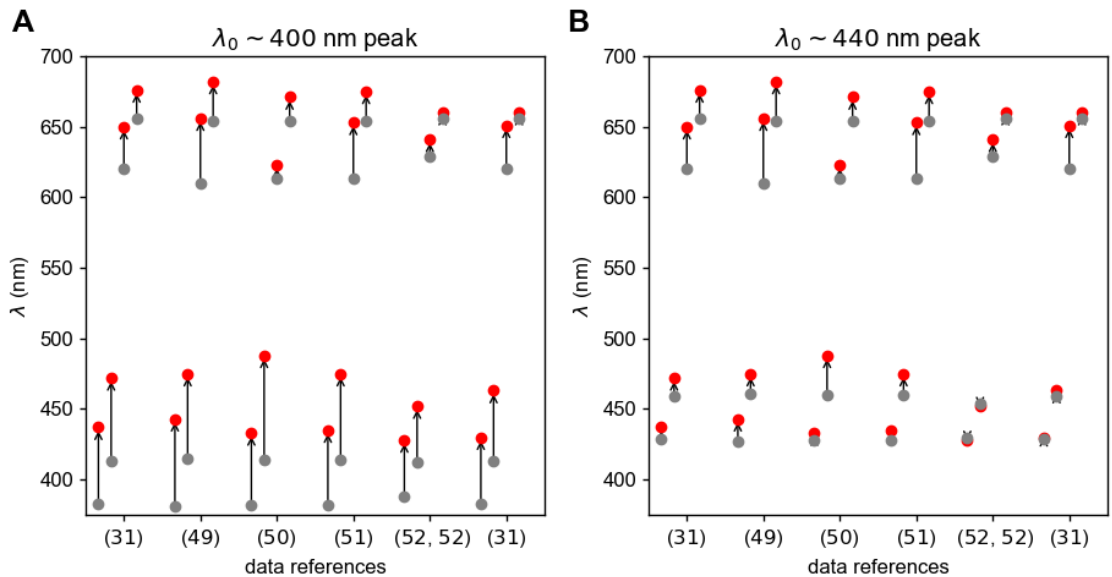

**Fig. S12**. (**A**) Comparison between the model prediction, grey points, and the absorption peaks values form the literature, red points, using the $\lambda_0 \sim 400$ nm peak. (**B**) Comparison between the model prediction, grey points, and the absorption peaks values form the literature, red points, using the $\lambda_0 \sim 440$ nm peak.

**Section S3: Solar Spectral Irradiance Data**

The direct solar spectrum used in our calculation is the National Renewable Energy Laboratory's (NREL) reference air mass 1.5 spectrum.[28] The NREL data includes three possible spectra based on different conditions. For the result show in Fig. 2G and S4 we use the ETR spectrum as it is not affected by atmospheric or geographic conditions, which may be variable between different locations at different times. The NREL atmospheric spectra were considered, however their effect does not change the overall result, rather they change the relative values of peaks in the optimization (see section S2.2 for details).

The solar spectrum under canopy (Fig. S13) was directly measured using a USB4000 Ocean Optics spectrometer. This setup records an integrated spectrum from 344 nm to 1039 nm. Integration times between 1 and 5 seconds were used. Initial measurements were taken under dense shaded canopy producing spectra such as that shown in Fig. S13C. Natural gaps in the foliage allowed unfiltered light to pass and generated spectra with large background noise. To mitigate this a variety of green leaves from trees native to North America were held together in a filter that was then applied to the spectrometer and pointed directly to the sun.

Fig. S13A shows the attenuation of the spectrum as a function of the number of leaves in the filter. The blue line is an average of single leaf measurements for a variety of species. The spectrum is strongly suppressed between 400 nm and 700 nm with increasing leaf count. With four leaves (red line in S10A) this portion of the solar spectrum is completely attenuated. Applying three stacked leaves replicated the spectrum under canopy accurately while avoiding fluctuations and background noise from foliage gaps and non-uniform canopy density.

Differing leaf species show only very minor variations in their transmitted solar spectrum. Fig. S13B shows a variety of single species leaf stacks. Single leaves display differences mainly around 550 nm. In three and four leaf stacks these differences become trivially small and the species type no longer has any meaningful effect on the spectrum. All measurements were taken on the University of California Riverside campus on a near cloudless day in summer as close to noon as possible.

The underwater spectrum was calculated by applying a spectrum attenuation equation[54] to the NREL standard solar spectrum. The solar spectral irradiance $I(\lambda, z)$ is given by

$$I(\lambda, z) = I(\lambda, 0) * e^{-\sigma(\lambda)z} \tag{22}$$

where $z$ is depth under sea water. $I(\lambda, z = 0)$ is the intensity of the spectrum as a function of wavelength at the surface and $\sigma(\lambda)$ is the absorption of water as a function of wavelength. After analyzing a variety of different data we settled on compiling together data from Buiteveld[29] and Kou[30] for our calculations. These two sets were the most recent, had the highest degree of agreement between them, and covered a wide wavelength range.

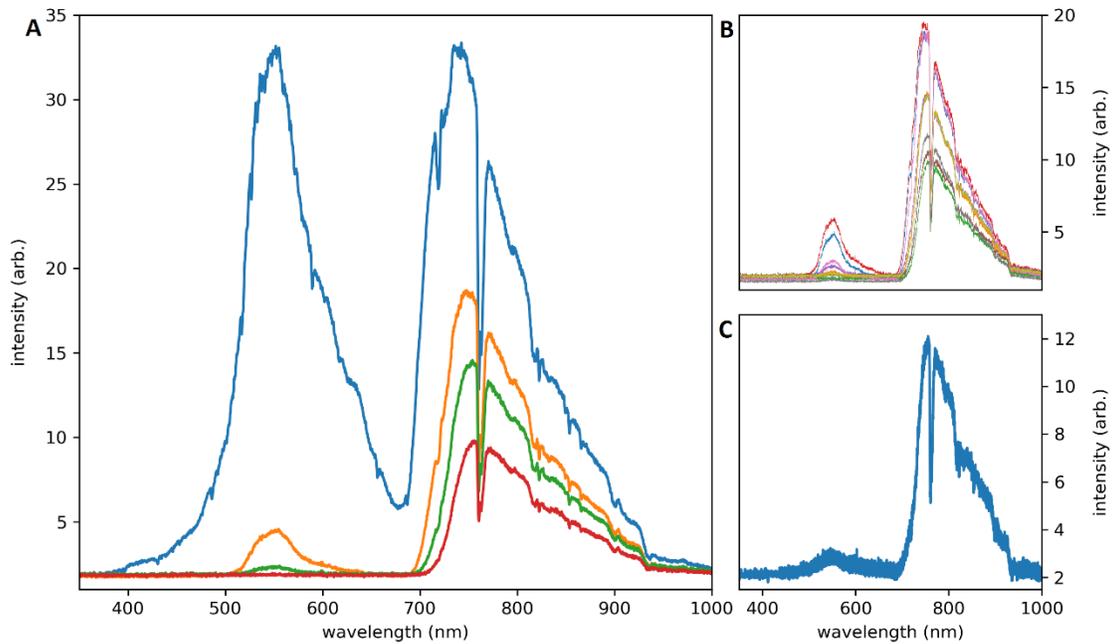

**Fig. S13.** (**A**), Average solar spectrum as measured through one, two, three, and four leaves of varying species (blue, orange, green, and red lines respectively). (**B**), Variance between species as leaf count increases, colored lines represent different leaf species. (**C**), Solar spectrum measured directly under canopy. The peak at 550nm fluctuates with time as a result of random gaps in the foliage.

**Section S4. Canopy Depth Dependence**

As discussed in the main text, we can test the noisy antenna model in systems where the solar spectrum is modulated as a function of some environmental parameter. While we consider sea water in the main text, we can also consider the depth within the canopy. In the case of purple bacteria, we consider the solar spectrum underneath a canopy of vegetation, but real canopies are not uniformly thick. The results presented in Fig. 2H of the main text are calculated from a spectrum measured deep under canopy, i.e. light has to pass through several leaves before it reaches the phototroph. In this case, the canopy cuts out almost all of the spectrum below 700 nm. Forests, however, are not always so dense, and detailed measurements[55] have shown variability with canopy density, composition, and conditions. The variability of forests, compared to the uniformity of seawater absorption, means that this test case is not as robust a test as underwater spectra, yet is still instructive.

To simulate the depth of the canopy we consider the number of leaves that light must pass through before it reaches the phototroph. Fig. S14A shows measured solar spectra under one to four leaves (see section S3 for details). Consistent with previous work[55], we see that light at wavelengths below 700 nm are strongly suppressed as a function of leaf number. But in the case of absorption by only one leaf (Fig. S14C), there would ideally be a set of peaks centered around 590 nm. For absorption by two leaves there might still be peaks near 600 nm, but they would absorb very little light. Under solar light suppression by three leaves there is no significant optimum at all below 700 nm. Thus, the model predicts that a phototroph that is always under minimal canopy would have a second absorber pair near 600 nm. While this is a contrived situation, it is a useful demonstration of the predictive elements of our model. Such predictions should be extended to other photosynthetic niches to test the detailed dependence of the absorption spectrum under particular environmental conditions.

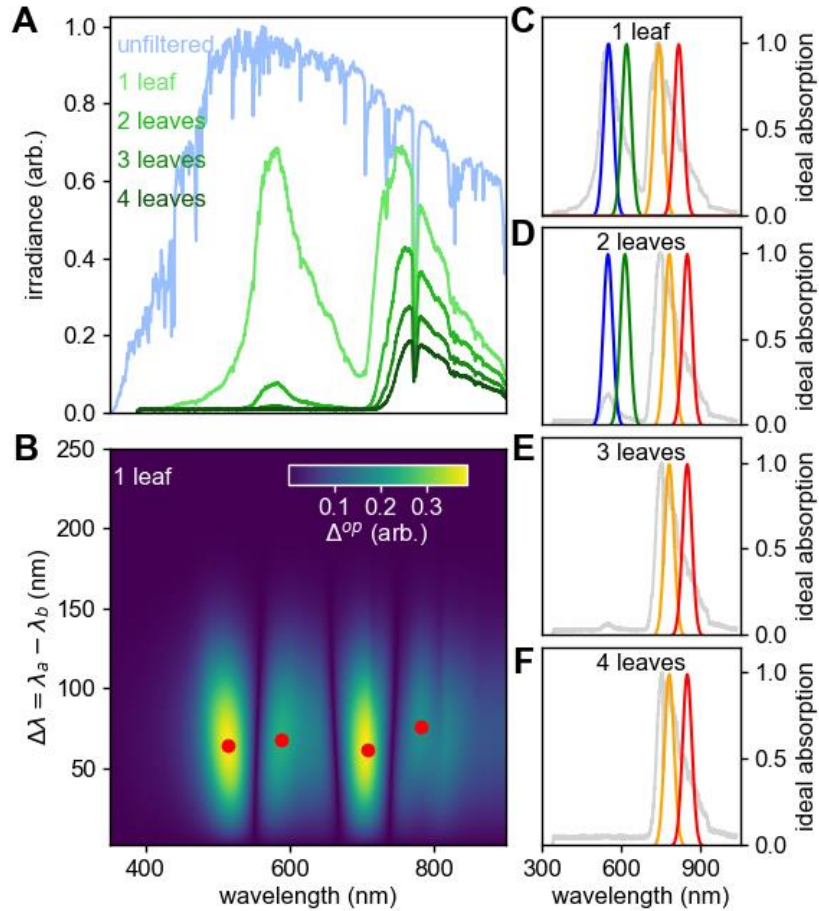

**Fig. S14**: The model calculation for spectra under a leaf canopy, adapted from main text Fig. 4. (**A**), The unfiltered solar spectrum and its attenuation by an integer number of leaves. (**B**), Calculation of $\Delta^{op}(\lambda_0, \Delta\lambda, w = 25 \text{ nm})$ versus center wavelength $\lambda_0$ and the peak separation $\Delta\lambda$ for the 1 leaf filtered solar spectrum. There are four maxima, red points, the first and third from the left are disallowed due to operable bandwidth considerations (see discussion in section S1.3). (**C-F**), Resulting allowed ideal absorption peaks predicted from this calculation for the spectra of different canopies.